\documentclass[aps,10pt,prd,letterpaper, amsmath, amssymb, preprintnumbers, showpacs, nofootinbib, superscriptaddress,tightenlines,twocolumn]{revtex4-1}
\usepackage{tikz}
\usepackage{mystyle_emacs}
\usepackage{stackrel}
\usepackage{theoremref}
\usepackage{placeins}
\usepackage{multirow}
\usepackage{graphicx}
\usepackage{xcolor}
\usepackage{float}

\usetikzlibrary{decorations.markings,decorations.pathmorphing}

\begin{document}

\title{Empirical investigation of nuclear correlation function distributions in lattice QCD}

\author{William Detmold}
\email{wdetmold@mit.edu}
\affiliation{Center for Theoretical Physics - a Leinweber Institute,
Massachusetts Institute of Technology, Cambridge, MA 02139, USA}
\author{Rohan Kanchana}
\email{rohankan@mit.edu}
\affiliation{Center for Theoretical Physics - a Leinweber Institute,
Massachusetts Institute of Technology, Cambridge, MA 02139, USA}
\author{Cagin Yunus}
\email{cyunus@mit.edu}
\affiliation{Center for Theoretical Physics - a Leinweber Institute,
Massachusetts Institute of Technology, Cambridge, MA 02139, USA}

\begin{abstract}
Two-point correlation functions of systems with baryon number $B\in\{1,2,3,4\}$ are investigated using lattice Quantum Chromodynamics (QCD). In particular, the empirical distributions of importance-sampling Monte-Carlo samples of these correlation functions are examined as a function of the spacetime separation between the two points and the baryon number. While the exact forms of these distributions are not known for QCD, recent work has determined asymptotic expressions for analogous correlation function distributions in simpler theories such as scalar field theory and the disordered phase of the $O(N)$ model. The theoretical $O(N)$ model distributions are found to provide an accurate description of the empirical QCD distributions at zero momentum over a wide range of temporal separations for each baryon number when assessed with a range of different statistical tests. In particular, the behaviour of the baryon number $B$ QCD correlation function at large temporal separation is well-reproduced by the  $O(N\sim 2/B)$ model distribution. 
\end{abstract}

\preprint{{MIT-CTP-5903}}
\maketitle

\newpage
\section{Introduction}

A direct understanding of the physics of nuclei from the underlying Standard Model of particle physics is an open problem. Nevertheless, it is widely anticipated that the nuclear physics emerges from the dynamics of quarks and gluons as governed by the theory of Quantum Chromodynamics (QCD). Since nuclei intrinsically depend on strong interactions at low energies compared to the QCD scale, $\Lambda_{\rm QCD}$, such an understanding requires calculations in the non-perturbative regime of QCD which thus far are only approachable with lattice field theory methods in the form of  lattice QCD (LQCD). 
This numerical approach relies on the Monte-Carlo sampling of configurations of the QCD degrees of freedom and is beset by a severe signal-to-noise problem when the quantities in question involve nuclei. Generalising heuristic arguments of Parisi \cite{Parisi:1983ae} and Lepage \cite{Lepage:1989hd}, Ref. \cite{Beane:2009gs} argued that for a nucleus of baryon number $B$, the ratio of signal-to-noise in a Euclidean two-point correlation function at zero three-momentum and at temporal separation $t$ is exponentially small:
\begin{equation}
    S/N \sim {\rm exp}\left( -B(m_N -3/2 m_\pi)t\right),
    \label{eq:stn}
\end{equation}
where $m_{N}$ and $m_{\pi}$ are the nucleon and pion masses, respectively. This scaling was seen to agree with the large $t$ behaviour of the numerical calculations in that and subsequent \cite{Beane:2009py,Beane:2010em,NPLQCD:2011naw,NPLQCD:2012mex,Orginos:2015aya,Wagman:2017tmp} work.
Finding a computational approach to problems involving nuclei that does not suffer from such a degradation of $S/N$ is an important goal of the field as a solution would allow calculations of a many physically interesting and phenomenologically important quantities that are as yet prohibitively difficult.

One avenue to pursue towards this goal is to investigate the empirical distributions found for QCD correlation functions. Previous work in this direction appears in Refs.~\cite{Beane:2009gs,Beane:2009py,Endres:2011mm,DeGrand:2012ik,Grabowska:2012ik,Wagman:2016bam,Wagman:2017gqi,Wagman:2017xfh,Davoudi:2020ngi}). 
Here, this is extended to multi-baryon systems and  the statistical distributions of LQCD two-point correlation functions for operators with the quantum numbers of the proton, deuteron, triton  and the $\alpha$ particle are analysed from the perspective of analytic distributions recently derived in the context of the $O(N)$ model \cite{Yunus:2022wuc,Yunus:2023dka}.  While LQCD involves very different degrees of freedom and interactions than the $O(N)$ model, remarkable agreement is found between the LQCD correlation functions and the $O(N)$ distributions at all time separations for each baryon number. The description by the $O(N)$ model distribution is considerably more accurate than other distributions proposed in the literature \cite{Wagman:2016bam,Wagman:2017gqi,Wagman:2017xfh}.

Intriguingly, the best-fit values of $N$ for the $O(N)$-model distributions are non-integer. At large time separations the fitted values of $N$ asymptote and scale as $N\sim2/B$, where $B$ is the baryon number of the system under consideration. The inverse correlation of these quantities is strikingly well-resolved and it is interesting to ask whether there is an underlying reason for it. Speculatively, the emergence of $O(N\to0)$ model dynamics in the limit of $B\to\infty$ is related to self-avoiding random walks \cite{deGennes:1972zz} in the context of the hopping expansion.

This work is set out in the following manner. Section \ref{sec:ON} reviews the probability distributions of correlation functions in the $O(N)$ model that were derived previously.
Section \ref{sec:data} first introduces the numerical LQCD correlation function used in this work before outlining the fitting procedures and fit results found in comparing the empirical LQCD data to the $O(N)$-model distributions. Finally, Sec.~\ref{sec:discussion} presents a broader perspective on the results and outlines possible future directions of investigation. The appendix presents more details of the fits that are obtained.

\section{$O(N)$ correlation function distributions}
\label{sec:ON}

The probability distribution of the simplest zero-momentum, $O(N)$-invariant two-point correlation function in the $O(N)$ model, $C(t)=\langle \sum_{a=1}^N \overline\phi_a(t) \overline\phi_a(0)\rangle$ (where $\overline\phi_a(t)$ is the $(a\in\{1,\ldots,N\}$)th spatially-averaged component of the $O(N)$ field),  was derived in Ref.~\cite{Yunus:2023dka} based on correlated mean-field arguments. 
On average $C(t)$ is real,  so the focus of this work is on the real part of the correlation function. The distribution of this real part is given by 
\begin{eqnarray}
\label{eq:shifted_func}
P(x;\omega_+,\o_-,N) 
&=& e^{\ff 1 2 (\o^--\o^+) x }\ff{(\o^+ \o^-)^{\ff N 2}}{(\o^+ + \o^-)^{\ff {N-1} 2}} \\
&&\times \ff{\abs{x}^{\ff{N-1}{2}}}{\ss{\pi}\Gamma(\ff N 2)}   K_{\ff{N-1}{2}} \lp \ff{1}{2}(\o^++\o^-)\abs{x}\rp, \nonumber
\end{eqnarray}
where $x=C(t)$ represents the value of the correlation function at time $t$ on a particular  $O(N)$ field-configuration drawn from the Boltzmann-weighted distribution. The parameters $\omega^\pm$ are determined by the interactions of the theory and $K_n(x)$ is a modified Bessel function of the second kind.

At large times, Ref.~\cite{Yunus:2023dka} argues that in the $O(N)$-model context, the two parameters $\o^\pm \to \bar\omega$  in Eq.~\eqref{eq:shifted_func}, so it follows that the simplified distribution
\begin{eqnarray}
P^{(0)}(t;\bar\omega,N) &=& \lim_{t\to \ii} P(x;\bar\omega,\bar\omega,N)  \nonumber \\
&=& \ff{2^{\ff{1-N}{2}}\bar\omega}{\Gamma \lp \ff N 2 \rp \ss \pi}(\bar\omega \abs{x})^{\ff{N-1}{2}}K_{\ff{N-1}{2}}\lp \bar\omega\abs{x} \rp
\label{eq:asym_func}
\end{eqnarray}
can describe the correlation function distributions.

In what follows, Eq.~\eqref{eq:shifted_func} will be used to attempt to describe the empirical distributions of the LQCD correlation functions arising from the Monte-Carlo generation process. Since QCD is a very different theory to the $O(N)$ model, one should not necessarily expect such a description to work well. 
Nevertheless, as will be discussed below, the $O(N)$ form is seen to describe LQCD correlation-function distributions remarkably well. 

\section{Empirical Investigations}

\subsection{QCD data set}
\label{sec:data}

The data considered here has been previously generated by the NPLQCD collaboration and is a subset of the data used in the studies of Ref.~\cite{Orginos:2015aya,Parreno:2021ovq}. In particular the correlation functions for the proton ($p$), deuteron ($d$), triton ($t$) and helium-4 ($\alpha$) computed on a single ensemble of gluon field configurations are used. As detailed in Ref.~\cite{Orginos:2015aya}, the ensemble was generated using the hybrid Monte-Carlo algorithm \cite{Duane:1987de} for tree-level-improved clover Wilson fermions \cite{Sheikholeslami:1985ij} with a L\"uscher-Weisz gauge action \cite{Luscher:1984xn} with one level of stout \cite{Morningstar:2003gk} smearing. The lattice volume was $L^3\times T=32^3\times96$ and the bare coupling and light (up and down) and strange quark mass parameters correspond to a lattice spacing of $a\sim 0.145$ fm,  a pion mass, $m_\pi\sim450$ MeV, and a kaon mass, $m_K\sim596$ MeV. In what follows, the lattice spacing is set to unity for simplicity, so all quantities are reported in their respective lattice units.

Zero-momentum correlation functions 
\begin{equation}
    C_B(t)= \sum_{\bf x} \langle \Omega | \tilde\chi_B({\bf x},t) \chi^\dagger_B({\bf 0},0)|\Omega\rangle 
\end{equation}
were computed for each spin state of the various nuclei and averaged with appropriate normalisations since the statistical behaviour of each spin state was seen to be consistent. Here, $|\Omega\rangle$ is the vacuum state and $\chi_B^\dagger({\bf x},t)$ and $\tilde\chi_B^\dagger({\bf x},t)$ are interpolating operators with the quantum numbers of the state $B\in\{p,d,t,\alpha\}$ built using the techniques introduced in Ref.~\cite{Detmold:2012eu}. As discussed in Refs.~\cite{Detmold:2012eu,Orginos:2015aya}, correlation functions were computed from gauge-covariantly smeared sources and for either point (SP) or smeared (SS) sink operators. Since the statistical behaviour of the SS and SP correlation function distributions is found to be similar, only the SP results are shown in this work, and both the forward and backward propagating signals are combined with appropriate parity projection. A total of $N_{\rm cfg}=1847$ configurations are studied with an average of $\overline{N}_{\rm src}=55$ randomly chosen source locations per configuration (these source locations are not averaged over in the analysis). 

\begin{figure}[!ht]
    \centering
\hspace*{-1.2cm}\includegraphics[width=1.2\columnwidth]{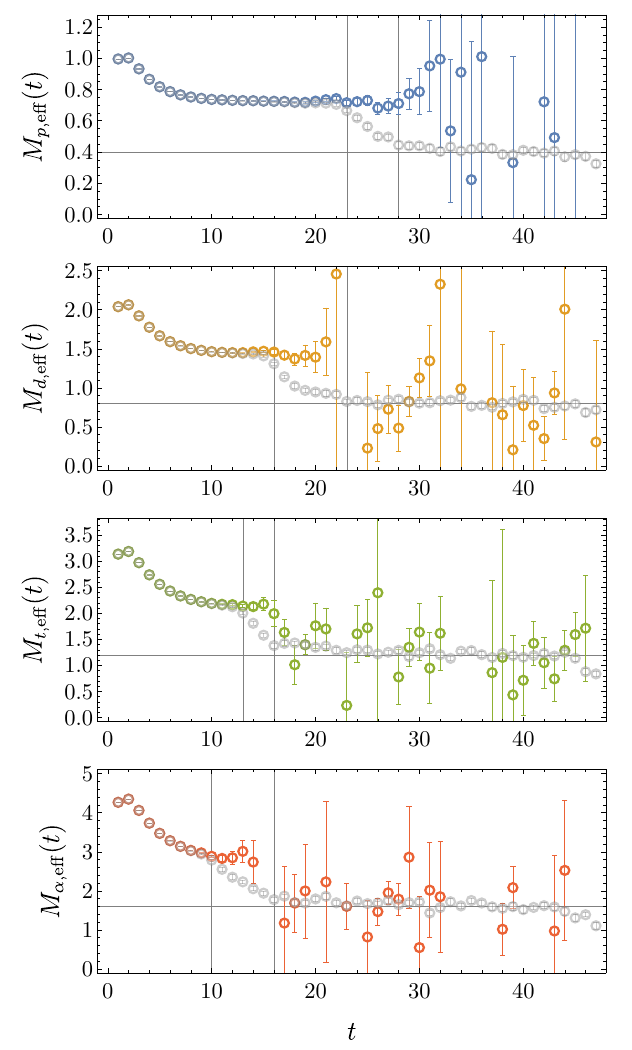}
    \caption{The effective mass functions for the various systems studied here (coloured, only shown when the mean of the correlation function is positive). The effective mass functions of the absolute values of the correlation functions are also shown in grey. The horizontal line in each panel indicates the scale of $3m_\pi B/2$, where $B$ is the baryon number of the system under consideration. The vertical lines indicate the time at which the effective mass and the effective mass of the absolute value of the correlation functions begin to deviate (${t}_{\rm noise}(B)$), and the time at which the latter effective mass becomes consistent with $3m_\pi B/2$ ($\tilde{t}_{\rm noise}(B)$).}
    \label{fig:effmass}
\end{figure}

The effective mass functions 
\begin{equation}
    M_{B,\rm eff}(t) = \ln \left[\frac{C_B(t)}{C_B(t+1)}\right]
\end{equation}
are shown for $B\in \{p,d,t,\alpha\}$ in Fig.~\ref{fig:effmass}. 
The uncertainties on the effective mass functions are computed using bootstrap resampling.
As can be seen, the signal-to-noise ratio  decreases approximately exponentially with $t$. To explore this, the effective mass functions of the absolute values of the correlation functions are also shown, following Ref.~\cite{Wagman:2016bam}. 
At early times, the correlation function and its absolute value agree, but eventually at $t_{\rm noise}(B)$, the phase of the correlation function becomes important and the uncertainties on the correlation function grow rapidly, as expected from Eq.~\eqref{eq:stn}. Empirically for this data set, 
\begin{equation}\begin{aligned}
t_{\rm noise}(p)\sim23, &
&t_{\rm noise}(d)\sim16, \\
t_{\rm noise}(t)\sim13, & &
t_{\rm noise}(\alpha)\sim10.     
\end{aligned}
\label{eq:tnoise}
\end{equation}
At large times, $t\agt \tilde{t}_{\rm noise}(B)$, the effective mass of the absolute value of the correlator is seen to saturate at the scale $3m_\pi B/2$, where $B$ is the baryon number of the system under consideration. For each system, 
\begin{equation}\begin{aligned}
\tilde t_{\rm noise}(p)\sim28, &
&\tilde t_{\rm noise}(d)\sim22, \\
\tilde t_{\rm noise}(t)\sim16, & &
\tilde t_{\rm noise}(\alpha)\sim16.     
\end{aligned}
\label{eq:tildetnoise}
\end{equation}

\subsection{Fitting methodology}
\label{sec:fitting}

For each system (labelled here as $B$) and each choice of $t$, the  procedure described in this subsection is used to find the best-fit $O(N)$-model parameters and uncertainties.

\begin{enumerate}
    \item Forward and backward correlation functions are concatenated to produce a set of samples $\{C_i\}$ with $i\in\{1,\ldots,2N_{\rm cfg}\overline N_{\rm src}\}$ indexing both the configuration and source location (appropriately translated) used for the correlation function. 
    For each timeslice, the data are ordered and truncated by removing the lowest and highest $\Delta_{\rm cut}=10\%$ in order to avoid numerical issues in subsequent analysis. The remaining data range between $C_-=\min_i \{C_i\}$ and $C_+=\max_i\{C_i\}$, evaluated on the truncated data. Using $N_{\rm bins}= 1600$ equally spaced bins of size $s = (C_+ - C_-) / N_{\rm bins}$, a density histogram $E_{B,t}(x)$ is produced for a discrete set $x_j = C_- + s (j - \frac12)$ where $j \in \{1, \ldots, N_{\rm bins} \}$.\footnote{Variations of the cuts used in data truncation and the number of bins used to produce the histograms have been investigated. For variations around the values of $\Delta_{\rm cut}$ and $N_{\rm bins}$ chosen here, the analysis is robust provided that $N_{\rm bins}$ increases as $\Delta_{\rm cut}$ decreases (since the binning is uniform, this approximately maintains a constant resolution over the bulk of the distribution). At large times, using $\Delta_{\text{cut}}\leq1\%$ leads to numerical precision issues in evaluating the model function.}
 
    \item Each empirical distribution $E_{B, t}(x)$ is fit by minimising the $L_1$-norm (total variation)\footnote{Using the $L_1$-norm allows the empirical data to determine the  regions of the distribution that most constrain the fit.}
    \begin{equation}
    \label{eq:loss} 
        L_{B,t}(\theta,\alpha) = \sum_i 
        | E_{B,t}(x_i) - \alpha\ M_{B,t}(x_i,\theta)|,
    \end{equation}
    where $M_{B,t}(x,\theta)\equiv P (x; \omega^+, \omega^-, N ) $ is  the asymmetric  $O(N)$-model distribution in Eq.~\eqref{eq:shifted_func} with parameters $\theta=\{N,\omega^+,\omega^-\}$ (since the normalisation of the histogram changes with the number of bins and data cut, the normalisation $\alpha$ is also determined in the fit). 
    Fits are performed using the Nelder-Mead algorithm for each $B$ by sweeping through the $t$-values in increasing order, with parameters from $t-1$ used to set the starting search range for the parameters at $t$. In particular, for each parameter $p\in\{\alpha,N,\omega^+,\omega^-\}$, the starting search range is  $[r p_{t-1},r^{-1} p_{t-1}]$ where $p_{t-1}$ is the value of $p$ from timeslice $t-1$ and $r$ is a random number drawn from the uniform distribution on $[0.7,1.0]$. For $t$=1 the initial parameters are chosen randomly. 
    
    \item This procedure is performed for each of the $N_{\rm boot}=14$ bootstrap resamplings of the original data and the resulting bootstrap distributions are used to estimate the parameter and distribution uncertainties.\footnote{Although it is typical to use more than 14 bootstrap samples, a very large number of numerically difficult fits are performed below, so a relatively small number of resamplings are used. The effect of the number of bootstraps on the analysis presented here has been investigated by doubling the number of bootstrap ensembles, with differences found to be significantly less than other sources of uncertainty. } The bin ranges defined by the original data set are used for each bootstrap.
    In particular, the envelope of the middle 67\% quantile of the bootstrap samplings is shown as an uncertainty of the fitted function. 
    
    \item Since the loss landscape defined by Eq.~\eqref{eq:loss} is complicated, the search is repeated $N_{\rm iter}=10$ times, with the best-fit parameter values for each $B$, $t$ and data bootstrap updated if the loss decreases.
\end{enumerate}

\begin{figure*}
    \centering
    \includegraphics[width=0.83\textwidth]{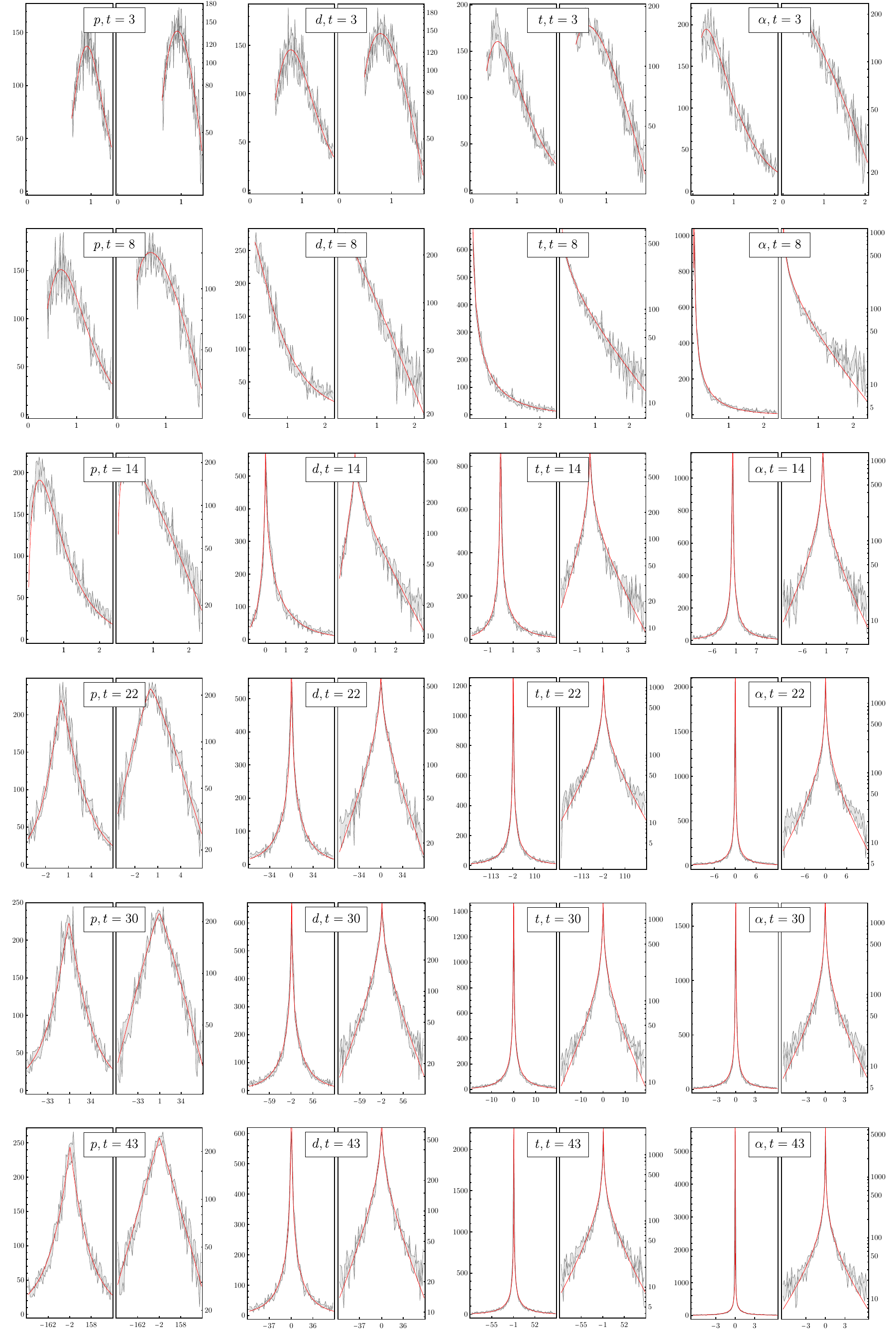}
    \caption{Representative histograms for each $B\in\{p,d,t,\alpha\}$ (left to right) and for $t\in\{3,8,14,22,30,43\}$. The grey regions show the bootstrap variation of the empirical histograms and the red curves are the best fits in each case.
    \label{fig:histogram_summary}}
\end{figure*}

\begin{figure}
    \centering
\includegraphics[width=\columnwidth]{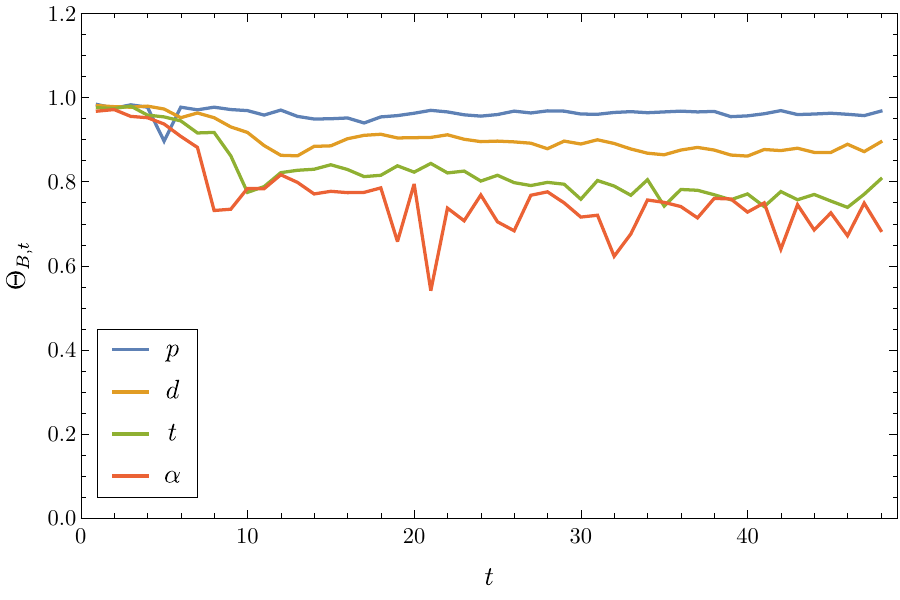}
\includegraphics[width=\columnwidth]{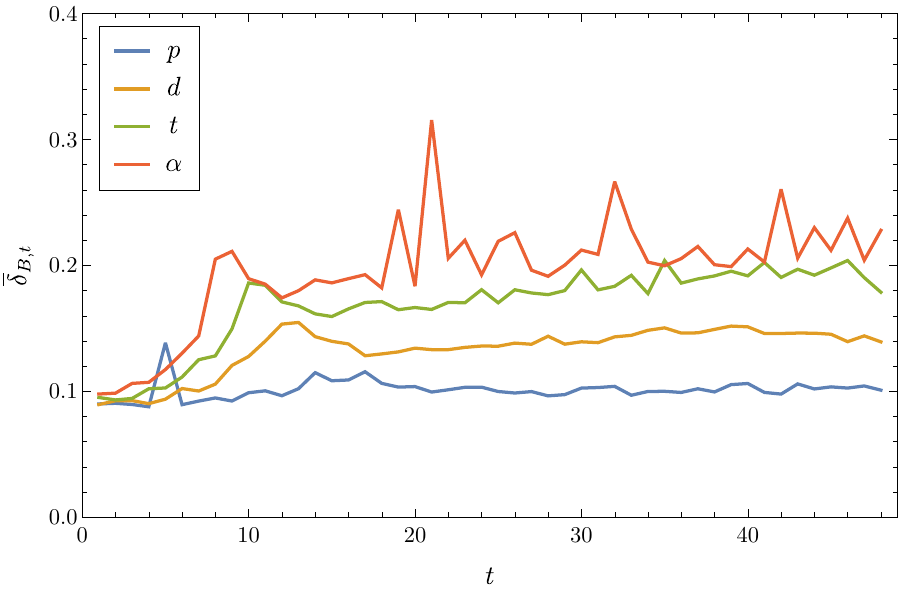}
    \caption{Upper panel: Fraction of bins for which the normalised fit residual $\delta_{B,t}(x)<0.3$ for each time and $B$. 
    Lower panel: Normalised fit residuals averaged over bins as a function of $t$ for each $B$. 
    \label{fig:fitresidual_summary}}
\end{figure}

\begin{figure*}
    \centering
\includegraphics[width=0.99\textwidth]{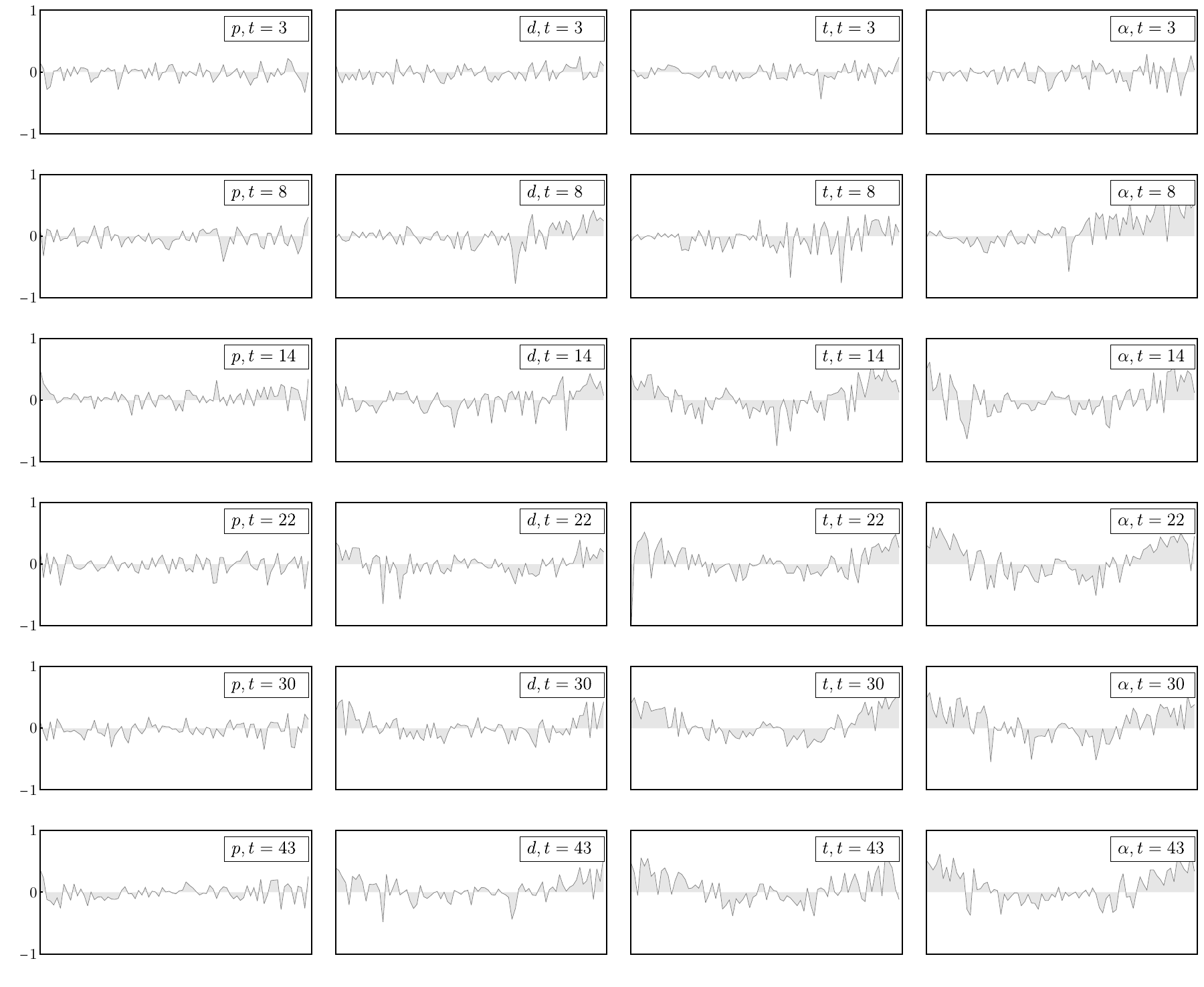}
    \caption{Example normalised fit residuals $\delta_{B,t}(x)$ for different time separations, $t\in\{3, 8, 14, 22, 30,43\}$ and baryon numbers, $B\in\{p,d,t,\alpha\}$. Residuals are shown for every 20th bin over the entire data range for each $B$ and $t$. 
    \label{fig:fitresidual_examples}}
\end{figure*}

The resulting fits are shown for all $B$ and for exemplary values of $t$ in Fig.~\ref{fig:histogram_summary}, with the full set of results shown in Figs. \ref{fig:dist_1}--\ref{fig:dist_8} in Appendix \ref{app:plots}. The extracted best fit parameters $\overline\theta=(N,\omega^+,\omega^-)$ are presented in Tables \ref{tab:fit_params_p}--\ref{tab:fit_params_a}  in Appendix \ref{app:plots} (the best fit normalisation, $\overline\alpha$, is not a free parameter but is set such that the model and empirical data are normalised equivalently). 
The description of the data for each $(B,t)$ combination has $|\overline\alpha\ M_{B,t}(x,\overline\theta)- E_{B,t}(x)|<\delta_{B,t}(x) |E_{B,t}(x)|$, with $|\delta_{B,t}(x)|<0.3$  in most  bins.\footnote{Because of the integrable singularity in the bin containing zero, that bin typically has a larger value of $\delta_{B,t}$ than other bins.} The fraction of bins with residuals less than 0.3 for a given $(B,t)$, $\Theta_{B,t}$ is shown in Fig.~\ref{fig:fitresidual_summary}. The average over all bins, $\overline{\delta}_{B,t}=\frac{1}{N_{\rm bins}}\sum_i |\delta_{B,t}(x_i)|$ is also shown for each $B$ as a function of $t$ in Fig.~\ref{fig:fitresidual_summary}. Figure~\ref{fig:fitresidual_examples} shows the normalised fit residuals, $\delta_{B,t}(x)$ for all $B$ and for the same exemplary values of $t$ as shown in Fig.~\ref{fig:histogram_summary}.
As can be seen from the figures, the overall description of the QCD data is remarkably accurate; the fitted distributions capture both the shape and magnitude of the empirical distributions, apart from occasionally overestimating the strength of the distribution in the zero-bin and underestimating the tails of the distribution at large $t$. The fits presented here are considerably more faithful in their representation of the empirical distributions than other model distributions that have been proposed previously for QCD correlation functions \cite{Wagman:2016bam,Wagman:2017gqi}.

\subsection{Temporal-separation and baryon-number dependence of fitted parameters}

\begin{figure}    
\includegraphics[width=\columnwidth]{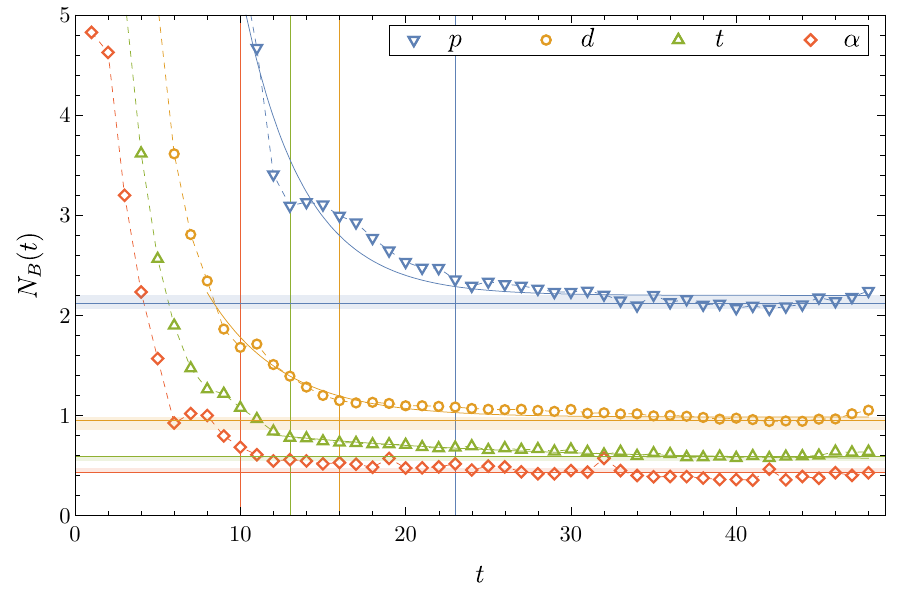}
\includegraphics[width=\columnwidth]{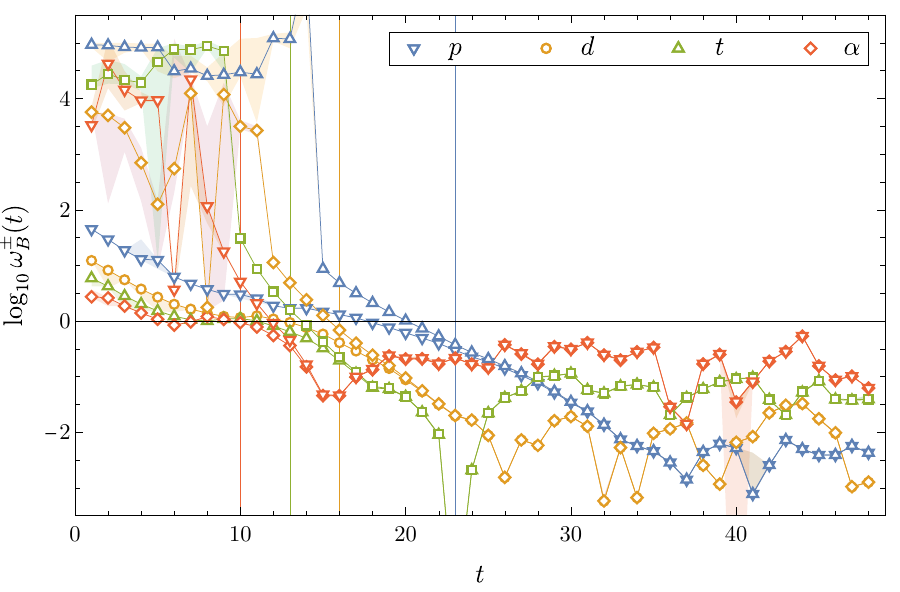}
\caption{Parameters extracted from correlation function fits for nucleon, deuteron, triton and $\alpha$. The lower panel shows the extracted values of $\log_{10}\omega^\pm$, while the upper panel shows the  extracted values of $N$ in Eq.~\eqref{eq:shifted_func} for each $B$ and $t$.
The displayed uncertainties are determined by the bootstrap procedure.
The extracted ranges of $N_B(t\to\infty)$ are shown as the horizontal bands, as discussed in the main text. Additionally, the accepted exponential fit with the smallest $t_{\text{min}}$ is shown as the solid line for each $B\in\{p,d,t\}$. In both panels, $t_{\rm noise}(B)$ is indicated for each $B$ by the vertical lines, increasing in $B$ from right to left.}
\label{fig:extracted_parameters}
\end{figure}

Having determined the best fit model parameters $\overline\theta$ for each $(B,t)$ pair, their dependence on $B$ and $t$ is investigated. Figure \ref{fig:extracted_parameters}, shows the extracted values of the parameters $N$ and $\omega_\pm$ for the three-parameter distribution in Eq.~\eqref{eq:shifted_func} as a function of $t$ for each $B$, defining $N_B(t)$ and $\omega_B^\pm(t)$, respectively. 

The extracted $N_B(t)$ is well-constrained in the fits and approximately saturates to a constant at large $t$ for each state, $B$.
The time extent at which this saturation occurs decreases as $B$ increases.
At early times, asymmetric fits with $\omega^-\ne\omega^+$ are preferred, but as $t$ increases, $\omega^\pm$ move towards each other, coinciding within uncertainties for $t\agt t_{\rm noise}(B)$. The $\omega^-$ parameter is numerically difficult to determine at small $t$ as large values are favoured in the fit and it appears inside exponentially growing functions.
At late times where $\omega^-\simeq\omega^+=\bar\omega$, the determination of $\bar\omega$ also becomes numerically challenging as the distributions become exponentially broader and the empirical data does not sufficiently sample the tails of the distribution, leading to the larger fluctuations with $t$ seen in Fig.~\ref{fig:extracted_parameters}.

For $B\in\{p,d,t\}$, the $O(N)$-model parameter $N_B(t)$ is well described by the model
\begin{equation}
    N_B(t) = {\cal N}_B + {\cal C}_B\ e^{-{\cal M}_B t}
    \label{eq:fitNvsT}
\end{equation}
over a large range of $t$ for each $B$, and by a constant at large $t$. To extract the asymptotic value, ${\cal N}_B=N_B(t\to\infty)$, fits to the above exponential model and to a constant (i.e., Eq.~\eqref{eq:fitNvsT} with ${\cal C}_B=0$) are performed over time ranges $t\in\{t_{\rm min},t_{\rm max}\}$ with $t_{\rm max}=48$ and varying $t_{\rm min}\in\{1,\ldots,40\}$ using $\chi^2$-minimisation. The envelope of the returned values of ${\cal N}_B$ for all fits that have a $\chi^2$ per degree of freedom less than $\chi^2_{\text{max}}/\text{dof}=1.8$ are then used to define the extracted uncertainty on ${\cal N}_B$ and the 50\% quantile is used for the central value. The mass scale ${\cal M}_B$, and in particular the coefficient ${\cal C}_B$, are less well determined. 
For $B=\alpha$, the saturation to a constant is less clear and there are significant fluctuations as $t$ varies.\footnote{This behaviour is likely due to the choice of $\delta_{\text{cut}}=0.1$ and $N_{\text{bins}}=1600$ in this analysis. A larger number of bins, or a different binning scheme, may give a better representation of the highly peaked data distribution the $B=\alpha$ data set at late times, but fits become computationally challenging in this case.} 
Consequently, the confidence interval of a constant fit over $t\in[25,48]$ is used for the central value and uncertainties in this case. Fig.~\ref{fig:extracted_parameters} shows the extracted ranges of ${\cal N}_B$ as shaded bands. Table \ref{tab:NBvsTfits} reports the fit parameters. Statistical errors on the parameters from a given fit are small compared to the variations between fits.

\begin{table}[!t]
    \centering
    \begin{ruledtabular}
    \begin{tabular}{cccc}
       $B$  & ${\cal N}_B$ & ${\cal C}_B$ & ${\cal M}_B$ \\ \hline
 $p$ &  $2.122_{-0.012}^{+0.027}$ & $5.5_{-2.9}^{+171.}$ & $0.16_{-0.04}^{+0.15}$ \\
 $d$ & $0.956_{-0.06}^{+0.020}$ & $1.8_{-1.1}^{+633.}$ & $0.12_{-0.08}^{+0.16}$ \\
 $t$ & $0.588_{-0.031}^{+0.001}$ & $1.8_{-1.0}^{+2200}$ & $0.09_{-0.04}^{+0.24}$ \\
 $\alpha$ & $0.433_{-0.011}^{+0.04}$ & -- & -- \\
            \end{tabular}
    \end{ruledtabular}
    \caption{Parameters determined by fits to the extracted $N_B(t)$ using Eq.~\eqref{eq:fitNvsT}. The central values are given by the 50\% quantile of the distribution of the parameters for all successful fits ($\chi^2/\text{dof} < \chi^2_{\text{max}}/\text{dof}$) and the uncertainties are from the 17\% and 83\% quantiles. For $B=\alpha$, only constant fits are used. 
    \label{tab:NBvsTfits}}
\end{table}

\begin{figure}
    \centering
    \includegraphics[width=\columnwidth]{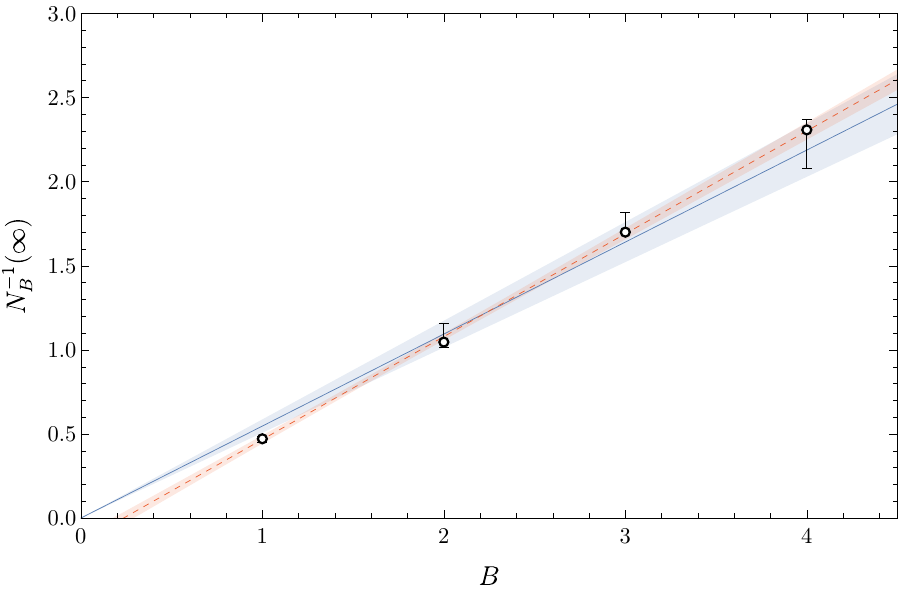} 
    \caption{The asymptotic value of $N_B^{-1}$ at large $t$ as a function of baryon number.
    Fits of the form $N^{-1}_B(\infty)= c_0+c_1 B$ are shown for $c_0$ unconstrained (red dashed line and surrounding uncertainty band) and for $c_0=0$ (blue solid line and surrounding uncertainty band). 
    \label{fig:paramsvsB}}
\end{figure}

Some interesting conclusions can be drawn from these fits.
Fig.~\ref{fig:paramsvsB} shows the extracted value of $N_B^{-1}(t\to\infty)$ as a function of the baryon number $B$. Also shown are fits to these results using the form $N^{-1}_B(\infty)= c_0+c_1 B$ with $c_0=0$ (linear) and $c_0$ free (linear plus constant). The linear fit  results in $c_1=0.55^{+0.02}_{-0.02}$, while the linear plus constant fit results in $c_0=-0.14^{+0.02}_{-0.02}$ and $c_1=0.61^{+0.01}_{-0.01}$, corresponding to a zero-crossing at $B\simeq0.24$. Qualitatively that suggests the intriguing scaling, $N_B(\infty)\sim 2/B$, which will be discussed further below.
A second observation is that the deviations from the asymptotic behaviour for $t\alt t_{\rm noise}(B)$ change the value of $N$ in an exponential manner with an exponent (${\cal M}_B$) that is commensurate with $m_\pi$ or with the QCD scale $\Lambda_{\text{QCD}}$. 

\subsection{Phase distributions of correlation functions}

\begin{figure}[!h]
    \centering
    \includegraphics[width=\columnwidth]{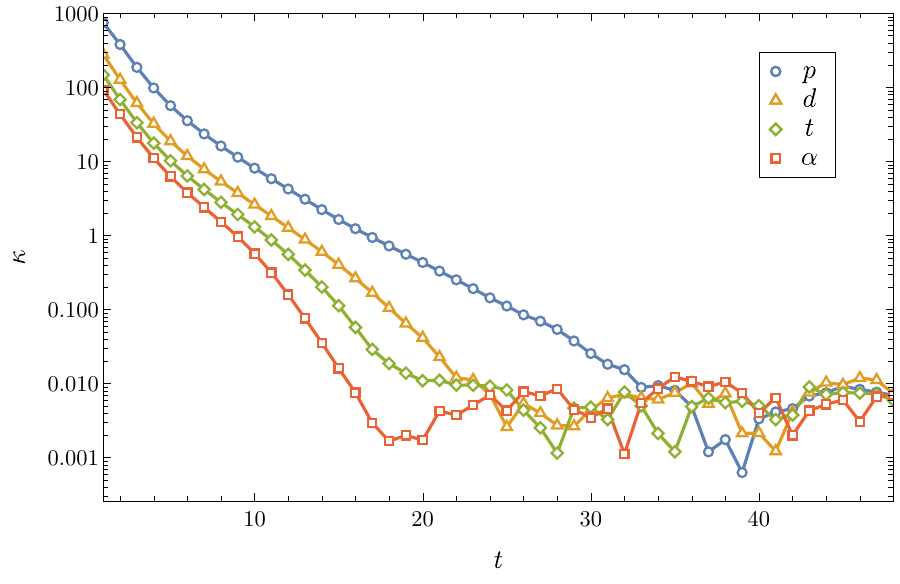}
    \caption{The von Mises concentration parameter $\kappa$ from fits to the phase distributions of the correlation functions to Eq.~\eqref{eq:vonMises} for each system.}
    \label{fig:concentration}
\end{figure}

\begin{figure*}[!t]
    \centering
\includegraphics[width=0.85\textwidth]{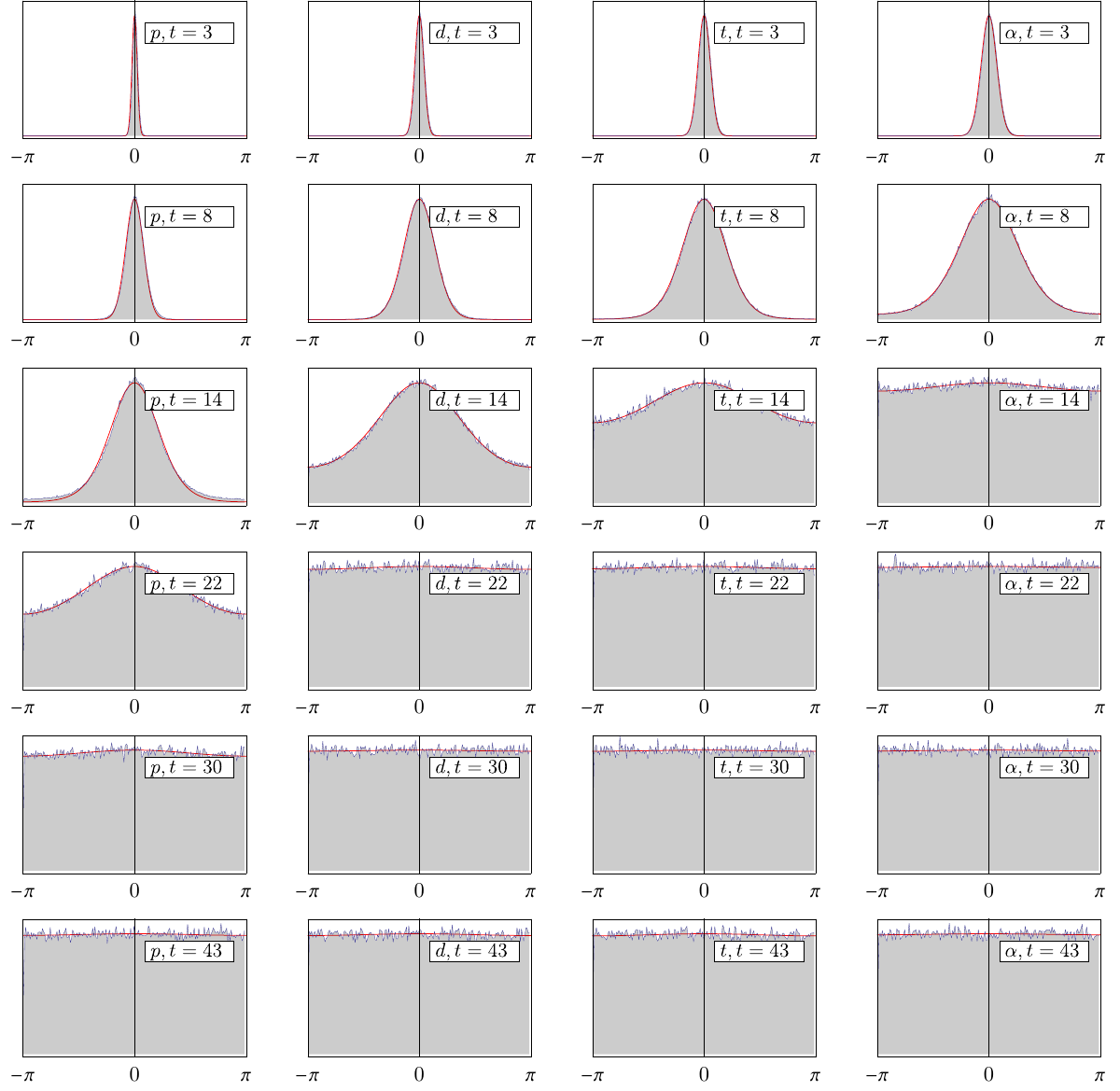}
    \caption{Phase distributions for different time separations, $t\in\{3, 8, 14, 22, 30,43\}$ and baryon numbers, $B\in\{p,d,t,\alpha\}$. The grey histogram represents the empirical distribution while the red curve is a fit using Eq.~\eqref{eq:vonMises}.}
    \label{fig:phase-deuteron}
\end{figure*}

In the above analysis, the focus has been on the real part of the correlation functions. It is also instructive to consider the distribution of the phase of the various correlation functions. While in both the $O(N)$ model and LQCD, the correlation functions investigated here are real on ensemble average, their expectation values arise from complex-valued distributions.  
Previous studies \cite{Wagman:2016bam} have found that QCD baryon phase distributions are described by wrapped-normal (von Mises) distributions
\begin{equation}
    f(x; \mu ,\kappa )=\frac {\exp{\kappa \cos{x-\mu }}}{2\pi I_{0}(\kappa )},
    \label{eq:vonMises}
\end{equation}
where $x\in[-\pi,\pi)$, $\mu$ is the mean of the distribution,  $\kappa$ is the concentration parameter, and $I_0(x)$ is a modified Bessel function of the first kind.
Similar behaviour is seen in the numerical LQCD data studied here. For each $(B,t)$, fits to the von Mises distribution with $\mu=0$ are performed and provide a description of the distribution with a goodness-of-fit $R^2>0.99$ for all systems for all $t\ge4$  (and $R^2>0.8$ for all $t$).
Figure~\ref{fig:concentration} shows the concentration parameters extracted in these fits, showing that for each system the concentration decreases as $t$ increases, eventually saturating at a very small value below which there is insufficient statistical power to cleanly resolve the value. Examples of the phase distributions are shown in Fig.~\ref{fig:phase-deuteron}.

\section{Discussion}
\label{sec:discussion}
In this work, the probability distribution of the simplest zero-momentum, $O(N)$-invariant two-point correlation function of the $O(N)$ model has been seen to provide an accurate description of the empirical distributions of $N_f=2+1$ flavour QCD correlation functions of baryon number $B\in\{1,2,3,4\}$ at all temporal separations, $t$. The QCD calculations were performed at quark masses corresponding to a pion mass of $m_\pi\sim450$~MeV and kaon mass of $m_K\sim 596$~MeV for one lattice spacing and volume. The model distribution captures all of the qualitative features of the QCD data and agrees quantitatively with a normalised average residual smaller than 0.2 for almost all of the $(B,t)$ combinations. 

A key finding of this work is that the values of the parameter $N$ for which the QCD distributions are best fit are seen to asymptote at large time separations to a non-integer value $N_B(t\to\infty)\sim 2/B$, scaling inversely with the baryon number. If this scaling were to extend to larger baryon-number systems, it would imply the QCD empirical distributions for large nuclear systems follow those of the $O(N\to0)$ model. While the origin of such behaviour is unknown and may simply be a coincidence, it is intriguing. 
It is possible that there is some sense of universality in asymptotic correlation function distributions. The methods used to compute the $O(N)$-model distributions in Ref.~\cite{Yunus:2023dka} were relatively theory-agnostic and relied on mean-field arguments, and  similar behaviour could potentially be derived in QCD (although the presence of fermions significantly complicates this). Such a derivation could also allow improved estimators for the expectation values of the path integrals that the LQCD correlation functions correspond to \cite{Yunus:2023dka}. This relation would explain the similarity of the  distributions seen empirically in LQCD to those in $O(N)$ models, but would not explain the correlation between $B$ and $N$.
One speculative direction that may provide an explanation of this latter feature is from the connection of the $O(N\to0)$ model to self-avoiding random walks \cite{deGennes:1972zz}. A hopping-parameter expansion of the QCD $B$-baryon correlation function involves at least $3B$ quark lines connecting the source and sink interpolating operators. Because of the fermionic nature of these lines, they avoid occupying the same links of the spacetime lattice. As $B$ increases, the number of such lines increases, potentially leading to behaviour akin to the self-avoiding walk.


An important extension of the current work would involve understanding how the results change as the parameters used in the LQCD calculations change, that is, the quark masses, lattice spacing and lattice volume. An investigation of mass dependence could clarify the origin of the scale ${\cal M}_B$ found in the fits to the time dependence of $N_B(t)$, and an investigation with changing geometry could shed light on the connection to self-avoiding walks. Given the intriguing dependence observed for $N_B(t\to\infty)\sim 2/B$, it would also be interesting to analyse distributions of correlation functions for mesons and for (multi-)baryon systems at other values of the number of colours. It would be similarly interesting to understand whether connections to $O(N)$ model distributions occur for other classes of LQCD correlation functions beyond two-point correlations. Finally, it is interesting to consider how the observed descriptions of LQCD correlation function distributions by $O(N)$-model distributions can be applied to ratios of correlation functions, as used to extract energy differences.



\acknowledgments
We thank Tom DeGrand, David B. Kaplan, Ethan Neil, Martin Savage, Phiala Shanahan, and Michael Wagman for interesting discussions and the NPLQCD collaboration for producing the correlation function data used in this study.
This research used resources of the Oak Ridge Leadership Computing Facility at the Oak Ridge National Laboratory, which is supported by the Office of Science of the U.S. Department of Energy under Contract number DE-AC05-00OR22725
and the resources of the National Energy Research Scientific Computing Center (NERSC), a Department of Energy Office of Science User Facility using NERSC award NP-ERCAPm747. The research reported in this work made use of computing facilities of the USQCD Collaboration, which are funded by the Office of Science of the U.S. Department of Energy, and used the MIT Physics Submit computing environment.
This work was supported in part by the U.S. Department of Energy, Office of Science under grant Contract Number DE-SC0011090, by the SciDAC5 award DE-SC0023116.

\bibliography{refs}

\newpage
\clearpage
\appendix
\section{Fitted distributions}
\label{app:plots}

\begin{table}[H]
        \begin{ruledtabular}
    \begin{tabular}{c ccc }
$t$ & $N$ & $\omega^+$ & $\omega^-$  \\ \hline 
1 & $88.7_{-1.0}^{+1.3}$ & $44.6_{-0.4}^{+0.8}$ & $\left(9.2_{-1.0}^{+0.8}\right)\times 10^4$ \\
2 & $57.1_{-32.}^{+1.1}$ & $28.9_{-16.}^{+0.5}$ & $\left(8.1_{-0.9}^{+1.7}\right)\times 10^4$ \\
3 & $36.8_{-11.}^{+0.9}$ & $18.6_{-6.}^{+0.5}$ & $\left(8.3_{-8.}^{+1.7}\right)\times 10^4$ \\
4 & $25.4_{-0.7}^{+33.}$ & $12.8_{-0.4}^{+17.}$ & $\left(8.2_{-0.7}^{+1.5}\right)\times 10^4$ \\
5 & $24.5_{-8.}^{+0.8}$ & $12.4_{-4.}^{+0.4}$ & $\left(8.0_{-5.}^{+1.6}\right)\times 10^4$ \\
6 & $12.17_{-0.11}^{+0.19}$ & $6.21_{-0.07}^{+0.10}$ & $\left(3.1_{-0.8}^{+2.6}\right)\times 10^4$ \\
7 & $9.05_{-0.23}^{+0.15}$ & $4.66_{-0.12}^{+0.10}$ & $\left(3.4_{-0.5}^{+2.4}\right)\times 10^4$ \\
8 & $7.04_{-0.11}^{+0.11}$ & $3.70_{-0.04}^{+0.06}$ & $\left(2.5_{-0.5}^{+1.1}\right)\times 10^4$ \\
9 & $5.70_{-0.08}^{+0.14}$ & $3.04_{-0.04}^{+0.06}$ & $\left(2.7_{-1.9}^{+4.}\right)\times 10^4$ \\
10 & $5.58_{-0.9}^{+0.17}$ & $2.99_{-0.5}^{+0.08}$ & $\left(3.0_{-0.5}^{+9.}\right)\times 10^4$ \\
11 & $4.7_{-0.7}^{+1.1}$ & $2.52_{-0.35}^{+0.5}$ & $\left(2.7_{-2.4}^{+9.}\right)\times 10^4$ \\
12 & $3.408_{-0.018}^{+0.04}$ & $1.872_{-0.022}^{+0.028}$ & $\left(1.21_{-0.14}^{+0.24}\right)\times 10^5$ \\
13 & $3.095_{-0.035}^{+0.020}$ & $1.686_{-0.028}^{+0.014}$ & $\left(1.2_{-0.4}^{+0.4}\right)\times 10^5$ \\
14 & $3.13_{-0.06}^{+0.30}$ & $1.698_{-0.033}^{+0.20}$ & $\left(2.5_{-2.1}^{+0.6}\right)\times 10^6$ \\
15 & $3.11_{-0.05}^{+0.04}$ & $1.469_{-0.010}^{+0.04}$ & $8.63_{-0.4}^{+0.16}$ \\
16 & $2.995_{-0.027}^{+0.06}$ & $1.301_{-0.017}^{+0.023}$ & $4.82_{-0.06}^{+0.05}$ \\
17 & $2.928_{-0.015}^{+0.10}$ & $1.135_{-0.017}^{+0.035}$ & $3.15_{-0.06}^{+0.05}$ \\
18 & $2.773_{-0.035}^{+0.04}$ & $0.926_{-0.012}^{+0.019}$ & $2.105_{-0.013}^{+0.008}$ \\
19 & $2.65_{-0.05}^{+0.09}$ & $0.762_{-0.019}^{+0.027}$ & $1.453_{-0.023}^{+0.018}$ \\
20 & $2.53_{-0.06}^{+0.06}$ & $0.611_{-0.015}^{+0.014}$ & $1.014_{-0.015}^{+0.028}$ \\
21 & $2.47_{-0.04}^{+0.04}$ & $0.493_{-0.007}^{+0.009}$ & $0.735_{-0.014}^{+0.006}$ \\
22 & $2.47_{-0.04}^{+0.05}$ & $0.394_{-0.004}^{+0.007}$ & $0.529_{-0.004}^{+0.012}$ \\
23 & $2.358_{-0.031}^{+0.04}$ & $0.297_{-0.007}^{+0.005}$ & $0.372_{-0.009}^{+0.009}$ \\
24 & $2.295_{-0.06}^{+0.029}$ & $0.227_{-0.004}^{+0.005}$ & $0.271_{-0.005}^{+0.006}$ \\
25 & $2.34_{-0.04}^{+0.06}$ & $0.1821_{-0.0035}^{+0.004}$ & $0.2064_{-0.0022}^{+0.005}$ \\
26 & $2.312_{-0.022}^{+0.11}$ & $0.1401_{-0.0013}^{+0.007}$ & $0.1543_{-0.0008}^{+0.007}$ \\
27 & $2.30_{-0.04}^{+0.07}$ & $0.1051_{-0.0023}^{+0.0024}$ & $0.1146_{-0.004}^{+0.0020}$ \\
28 & $2.265_{-0.027}^{+0.06}$ & $0.0759_{-0.0010}^{+0.0020}$ & $0.0810_{-0.0010}^{+0.0026}$ \\
29 & $2.232_{-0.034}^{+0.07}$ & $0.0519_{-0.0012}^{+0.0012}$ & $0.0538_{-0.0007}^{+0.0013}$ \\
30 & $2.23_{-0.04}^{+0.07}$ & $0.0339_{-0.0008}^{+0.0011}$ & $0.0351_{-0.0011}^{+0.0005}$ \\
31 & $2.25_{-0.04}^{+0.05}$ & $0.0234_{-0.0004}^{+0.0004}$ & $0.0239_{-0.0005}^{+0.0005}$ \\
32 & $2.21_{-0.04}^{+0.06}$ & $0.01359_{-0.0004}^{+0.00030}$ & $0.01366_{-0.0004}^{+0.00032}$ \\
33 & $2.12_{-0.07}^{+0.06}$ & $0.00733_{-0.00022}^{+0.00026}$ & $0.00733_{-0.00015}^{+0.00028}$ \\
34 & $2.098_{-0.024}^{+0.05}$ & $0.00562_{-0.00010}^{+0.00007}$ & $0.00566_{-0.00014}^{+0.00009}$ \\
35 & $2.20_{-0.06}^{+0.04}$ & $0.00452_{-0.00012}^{+0.00008}$ & $0.00457_{-0.00016}^{+0.00007}$ \\
36 & $2.13_{-0.09}^{+0.04}$ & $0.00284_{-0.00009}^{+0.00008}$ & $0.00286_{-0.00011}^{+0.00006}$ \\
37 & $2.15_{-0.09}^{+0.04}$ & $0.001381_{-0.00005}^{+0.000029}$ & $0.001399_{-0.00007}^{+0.000017}$ \\
38 & $2.107_{-0.012}^{+0.04}$ & $0.00440_{-0.00008}^{+0.00007}$ & $0.004417_{-0.000011}^{+0.00007}$ \\
39 & $2.115_{-0.05}^{+0.021}$ & $0.00617_{-0.00010}^{+0.00011}$ & $0.00625_{-0.00018}^{+0.00004}$ \\
40 & $2.072_{-0.020}^{+0.04}$ & $0.00517_{-0.00007}^{+0.00013}$ & $0.00519_{-0.00007}^{+0.00012}$ \\
41 & $2.1_{-0.6}^{+29.}$ & $0.00075_{-0.00025}^{+0.0035}$ & $0.00077_{-0.00026}^{+0.0035}$ \\
42 & $2.07_{-0.08}^{+0.04}$ & $0.002551_{-0.00011}^{+0.000034}$ & $0.002569_{-0.00013}^{+0.000029}$ \\
43 & $2.09_{-0.05}^{+0.05}$ & $0.00723_{-0.00021}^{+0.00012}$ & $0.00723_{-0.00020}^{+0.00012}$ \\
44 & $2.11_{-0.05}^{+0.06}$ & $0.00493_{-0.00009}^{+0.00018}$ & $0.00496_{-0.00010}^{+0.00015}$ \\
45 & $2.18_{-0.06}^{+0.04}$ & $0.00387_{-0.00010}^{+0.00006}$ & $0.00387_{-0.00006}^{+0.00007}$ \\
46 & $2.136_{-0.09}^{+0.023}$ & $0.003868_{-0.00013}^{+0.000033}$ & $0.003878_{-0.00014}^{+0.000032}$ \\
47 & $2.18_{-0.04}^{+0.09}$ & $0.00562_{-0.00011}^{+0.00019}$ & $0.00569_{-0.00015}^{+0.00014}$ \\
48 & $2.244_{-0.028}^{+0.06}$ & $0.004255_{-0.000033}^{+0.00013}$ & $0.004274_{-0.000035}^{+0.00016}$
    \end{tabular}
   \end{ruledtabular}
    \caption{Fit parameters for the proton.}
    \label{tab:fit_params_p}
\end{table}
\begin{table}[H]
        \begin{ruledtabular}
    \begin{tabular}{c ccc }
$t$ & $N$ & $\omega^+$ & $\omega^-$  \\ \hline 
1 & $2.49_{-0.15}^{+2.8}$ & $1.43_{-0.04}^{+1.3}$ & $\left(0.13_{-0.12}^{+6.}\right)\times 10^3$ \\
2 & $2.49_{-0.15}^{+5.}$ & $1.43_{-0.04}^{+2.5}$ & $\left(0.13_{-0.13}^{+5.}\right)\times 10^3$ \\
3 & $10.56_{-7.}^{+0.25}$ & $5.45_{-3.4}^{+0.12}$ & $\left(4.0_{-2.7}^{+8.}\right)\times 10^3$ \\
4 & $5.1_{-2.7}^{+2.2}$ & $2.7_{-1.3}^{+1.1}$ & $\left(0.15_{-0.15}^{+3.2}\right)\times 10^3$ \\
5 & $2.46_{-0.21}^{+2.7}$ & $1.42_{-0.07}^{+1.3}$ & $\left(0.009_{-0.006}^{+7.}\right)\times 10^3$ \\
6 & $3.61_{-0.9}^{+0.13}$ & $1.99_{-0.5}^{+0.07}$ & $\left(0.0055_{-0.0035}^{+1.2}\right)\times 10^5$ \\
7 & $2.807_{-0.07}^{+0.034}$ & $1.629_{-0.04}^{+0.020}$ & $\left(1.23_{-0.12}^{+1.2}\right)\times 10^4$ \\
8 & $2.34_{-0.12}^{+0.17}$ & $1.385_{-0.023}^{+0.05}$ & $\left(0.00175_{-0.00024}^{+3.2}\right)\times 10^3$ \\
9 & $1.863_{-0.11}^{+0.027}$ & $1.205_{-0.024}^{+0.031}$ & $\left(1.2_{-1.2}^{+0.7}\right)\times 10^4$ \\
10 & $1.680_{-0.024}^{+0.012}$ & $1.149_{-0.022}^{+0.019}$ & $3147._{-335.}^{+927.}$ \\
11 & $1.713_{-0.019}^{+0.024}$ & $1.233_{-0.023}^{+0.016}$ & $2652._{-418.}^{+501.}$ \\
12 & $1.509_{-0.017}^{+0.013}$ & $1.081_{-0.011}^{+0.013}$ & $11.22_{-0.23}^{+0.4}$ \\
13 & $1.393_{-0.012}^{+0.018}$ & $0.936_{-0.007}^{+0.021}$ & $4.87_{-0.10}^{+0.09}$ \\
14 & $1.284_{-0.008}^{+0.012}$ & $0.762_{-0.014}^{+0.013}$ & $2.396_{-0.015}^{+0.030}$ \\
15 & $1.200_{-0.012}^{+0.015}$ & $0.579_{-0.010}^{+0.009}$ & $1.263_{-0.014}^{+0.017}$ \\
16 & $1.148_{-0.027}^{+0.017}$ & $0.4057_{-0.0034}^{+0.010}$ & $0.685_{-0.009}^{+0.013}$ \\
17 & $1.126_{-0.017}^{+0.014}$ & $0.288_{-0.004}^{+0.007}$ & $0.398_{-0.008}^{+0.007}$ \\
18 & $1.132_{-0.005}^{+0.010}$ & $0.2000_{-0.0029}^{+0.0028}$ & $0.2419_{-0.0029}^{+0.005}$ \\
19 & $1.120_{-0.009}^{+0.018}$ & $0.1400_{-0.0019}^{+0.0035}$ & $0.1573_{-0.0034}^{+0.0023}$ \\
20 & $1.097_{-0.011}^{+0.009}$ & $0.0877_{-0.0006}^{+0.0018}$ & $0.0944_{-0.0007}^{+0.0010}$ \\
21 & $1.098_{-0.010}^{+0.026}$ & $0.05345_{-0.00032}^{+0.0018}$ & $0.0554_{-0.0010}^{+0.0018}$ \\
22 & $1.091_{-0.010}^{+0.014}$ & $0.03154_{-0.00024}^{+0.0007}$ & $0.03226_{-0.0006}^{+0.00024}$ \\
23 & $1.085_{-0.008}^{+0.013}$ & $0.01960_{-0.00010}^{+0.00027}$ & $0.01995_{-0.00010}^{+0.00016}$ \\
24 & $1.070_{-0.007}^{+0.016}$ & $0.01639_{-0.00014}^{+0.00020}$ & $0.01653_{-0.00019}^{+0.00024}$ \\
25 & $1.062_{-0.009}^{+0.006}$ & $0.00876_{-0.00013}^{+0.00011}$ & $0.00876_{-0.00006}^{+0.00011}$ \\
26 & $1.058_{-0.017}^{+0.025}$ & $0.001533_{-0.000017}^{+0.000034}$ & $0.001538_{-0.000022}^{+0.00004}$ \\
27 & $1.051_{-0.010}^{+0.023}$ & $0.00722_{-0.00020}^{+0.00011}$ & $0.00722_{-0.00016}^{+0.00011}$ \\
28 & $1.051_{-0.006}^{+0.022}$ & $0.00579_{-0.00006}^{+0.00012}$ & $0.00579_{-0.00004}^{+0.00012}$ \\
29 & $1.036_{-0.5}^{+0.022}$ & $0.01594_{-0.016}^{+0.00024}$ & $0.01594_{-0.010}^{+0.00030}$ \\
30 & $1.060_{-0.023}^{+0.013}$ & $0.01906_{-0.00031}^{+0.0004}$ & $0.01908_{-0.00025}^{+0.0004}$ \\
31 & $1.022_{-0.027}^{+0.007}$ & $0.01265_{-0.00028}^{+0.00025}$ & $0.01277_{-0.00033}^{+0.00016}$ \\
32 & $1.027_{-0.009}^{+0.18}$ & $0.000574_{-0.000017}^{+0.00012}$ & $0.000580_{-0.000018}^{+0.00013}$ \\
33 & $1.014_{-0.033}^{+0.021}$ & $0.00528_{-0.00016}^{+0.00016}$ & $0.00528_{-0.00016}^{+0.00016}$ \\
34 & $1.017_{-0.011}^{+0.023}$ & $0.000665_{-0.000007}^{+0.000012}$ & $0.000665_{-0.000007}^{+0.000012}$ \\
35 & $0.996_{-0.5}^{+0.019}$ & $0.00941_{-0.009}^{+0.00019}$ & $0.00946_{-0.006}^{+0.00016}$ \\
36 & $1.0006_{-0.022}^{+0.0028}$ & $0.01138_{-0.00028}^{+0.00011}$ & $0.01146_{-0.00032}^{+0.00013}$ \\
37 & $0.991_{-0.012}^{+0.007}$ & $0.01441_{-0.00017}^{+0.00024}$ & $0.01457_{-0.00016}^{+0.00034}$ \\
38 & $0.981_{-0.007}^{+0.009}$ & $0.002511_{-0.000032}^{+0.000028}$ & $0.002536_{-0.000005}^{+0.000032}$ \\
39 & $0.9635_{-0.0026}^{+0.012}$ & $0.001160_{-0.000016}^{+0.000014}$ & $0.001164_{-0.000018}^{+0.000011}$ \\
40 & $0.968_{-0.4}^{+0.010}$ & $0.00654_{-0.006}^{+0.00004}$ & $0.006549_{-0.004}^{+0.000029}$ \\
41 & $0.951_{-0.006}^{+0.012}$ & $0.00824_{-0.00007}^{+0.00025}$ & $0.00831_{-0.00009}^{+0.00020}$ \\
42 & $0.939_{-0.015}^{+0.006}$ & $0.02217_{-0.00031}^{+0.00028}$ & $0.02217_{-0.00018}^{+0.00028}$ \\
43 & $0.945_{-0.007}^{+0.012}$ & $0.02977_{-0.00031}^{+0.0004}$ & $0.03044_{-0.0004}^{+0.00018}$ \\
44 & $0.939_{-0.4}^{+0.012}$ & $0.0315_{-0.031}^{+0.0009}$ & $0.0327_{-0.020}^{+0.0007}$ \\
45 & $0.963_{-0.006}^{+0.014}$ & $0.01731_{-0.00022}^{+0.00019}$ & $0.01740_{-0.00030}^{+0.0004}$ \\
46 & $0.966_{-0.011}^{+0.017}$ & $0.00950_{-0.00014}^{+0.00029}$ & $0.00973_{-0.00019}^{+0.00013}$ \\
47 & $1.009_{-0.4}^{+0.018}$ & $0.001032_{-0.0006}^{+0.000005}$ & $0.001046_{-0.0006}^{+0.000016}$ \\
48 & $1.053_{-0.04}^{+0.018}$ & $0.001261_{-0.00006}^{+0.000035}$ & $0.001268_{-0.00005}^{+0.000031}$
    \end{tabular}
   \end{ruledtabular}
    \caption{Fit parameters for the deuteron.}
    \label{tab:fit_params_d}
\end{table}

\begin{table}[H]
        \begin{ruledtabular}
    \begin{tabular}{c ccc }
$t$ & $N$ & $\omega^+$ & $\omega^-$  \\ \hline 
1 & $11.15_{-3.3}^{+0.21}$ & $5.87_{-1.7}^{+0.12}$ & $\left(1.77_{-0.24}^{+2.1}\right)\times 10^4$ \\
2 & $7.84_{-0.08}^{+0.07}$ & $4.19_{-0.04}^{+0.07}$ & $\left(2.8_{-0.9}^{+2.2}\right)\times 10^4$ \\
3 & $5.19_{-2.0}^{+0.14}$ & $2.80_{-1.1}^{+0.07}$ & $\left(2.14_{-0.34}^{+2.0}\right)\times 10^4$ \\
4 & $3.62_{-1.0}^{+0.04}$ & $2.008_{-0.5}^{+0.023}$ & $\left(1.9_{-0.6}^{+0.6}\right)\times 10^4$ \\
5 & $2.56_{-0.04}^{+0.21}$ & $1.509_{-0.030}^{+0.06}$ & $\left(4.5_{-4.}^{+3.5}\right)\times 10^4$ \\
6 & $1.899_{-0.024}^{+0.013}$ & $1.215_{-0.030}^{+0.015}$ & $\left(7.6_{-2.2}^{+2.2}\right)\times 10^4$ \\
7 & $1.472_{-0.010}^{+0.04}$ & $1.047_{-0.016}^{+0.026}$ & $\left(7.5_{-5.}^{+1.0}\right)\times 10^4$ \\
8 & $1.262_{-0.017}^{+0.014}$ & $0.997_{-0.005}^{+0.014}$ & $\left(8.8_{-0.8}^{+2.2}\right)\times 10^4$ \\
9 & $1.217_{-0.010}^{+0.017}$ & $1.087_{-0.022}^{+0.023}$ & $\left(7.1_{-4.}^{+1.0}\right)\times 10^4$ \\
10 & $1.077_{-0.007}^{+0.009}$ & $1.100_{-0.017}^{+0.007}$ & $30.5_{-1.6}^{+0.9}$ \\
11 & $0.9641_{-0.0018}^{+0.005}$ & $1.006_{-0.016}^{+0.014}$ & $8.65_{-0.13}^{+0.32}$ \\
12 & $0.841_{-0.011}^{+0.012}$ & $0.808_{-0.014}^{+0.029}$ & $3.37_{-0.09}^{+0.04}$ \\
13 & $0.778_{-0.008}^{+0.009}$ & $0.638_{-0.007}^{+0.019}$ & $1.584_{-0.025}^{+0.032}$ \\
14 & $0.772_{-0.004}^{+0.010}$ & $0.486_{-0.004}^{+0.013}$ & $0.830_{-0.011}^{+0.008}$ \\
15 & $0.742_{-0.016}^{+0.011}$ & $0.321_{-0.007}^{+0.005}$ & $0.425_{-0.012}^{+0.011}$ \\
16 & $0.729_{-0.013}^{+0.011}$ & $0.1944_{-0.0023}^{+0.005}$ & $0.224_{-0.004}^{+0.004}$ \\
17 & $0.726_{-0.005}^{+0.009}$ & $0.1126_{-0.0021}^{+0.0016}$ & $0.1211_{-0.0019}^{+0.0018}$ \\
18 & $0.7124_{-0.0010}^{+0.0034}$ & $0.0629_{-0.0006}^{+0.0013}$ & $0.0664_{-0.0006}^{+0.0012}$ \\
19 & $0.713_{-0.006}^{+0.008}$ & $0.0597_{-0.0008}^{+0.0013}$ & $0.0611_{-0.0010}^{+0.0017}$ \\
20 & $0.705_{-0.018}^{+0.007}$ & $0.0427_{-0.0015}^{+0.0008}$ & $0.0432_{-0.0013}^{+0.0008}$ \\
21 & $0.664_{-0.33}^{+0.029}$ & $0.0215_{-0.021}^{+0.0018}$ & $0.0225_{-0.022}^{+0.0011}$ \\
22 & $0.667_{-0.34}^{+0.010}$ & $0.00892_{-0.009}^{+0.00020}$ & $0.00912_{-0.006}^{+0.00026}$ \\
23 & $0.680_{-0.007}^{+0.009}$ & $\left(4.23_{-0.04}^{+0.10}\right)\times 10^{-6}$ & $\left(4.23_{-0.04}^{+0.10}\right)\times 10^{-6}$ \\
24 & $0.693_{-0.011}^{+0.010}$ & $0.00207_{-0.00005}^{+0.00005}$ & $0.00207_{-0.00005}^{+0.00006}$ \\
25 & $0.655_{-0.006}^{+0.010}$ & $0.02198_{-0.00024}^{+0.0006}$ & $0.0223_{-0.0005}^{+0.0004}$ \\
26 & $0.672_{-0.24}^{+0.013}$ & $0.0410_{-0.04}^{+0.0011}$ & $0.0416_{-0.022}^{+0.0014}$ \\
27 & $0.653_{-0.24}^{+0.010}$ & $0.0539_{-0.05}^{+0.0009}$ & $0.0539_{-0.023}^{+0.0013}$ \\
28 & $0.656_{-0.24}^{+0.019}$ & $0.094_{-0.09}^{+0.007}$ & $0.096_{-0.05}^{+0.004}$ \\
29 & $0.637_{-0.34}^{+0.007}$ & $0.1022_{-0.10}^{+0.0027}$ & $0.1022_{-0.08}^{+0.0027}$ \\
30 & $0.6607_{-0.29}^{+0.0029}$ & $0.113315_{-0.11}^{+0.000009}$ & $0.1137_{-0.11}^{+0.0028}$ \\
31 & $0.636507_{-0.}^{+0.006}$ & $0.057441_{-0.}^{+0.00034}$ & $0.057441_{-0.}^{+0.00034}$ \\
32 & $0.611_{-0.22}^{+0.019}$ & $0.0498_{-0.05}^{+0.0026}$ & $0.0498_{-0.025}^{+0.0030}$ \\
33 & $0.621_{-0.21}^{+0.029}$ & $0.0648_{-0.06}^{+0.0030}$ & $0.0653_{-0.034}^{+0.0033}$ \\
34 & $0.596414_{-0.}^{+0.010}$ & $0.071926_{-0.}^{+0.0017}$ & $0.071926_{-0.}^{+0.0019}$ \\
35 & $0.622421_{-0.}^{+0.015}$ & $0.0638107_{-0.}^{+0.0008}$ & $0.064451_{-0.}^{+0.00012}$ \\
36 & $0.6154_{-0.0015}^{+0.005}$ & $0.02032_{-0.00006}^{+0.00010}$ & $0.02032_{-0.00006}^{+0.00010}$ \\
37 & $0.582_{-0.19}^{+0.010}$ & $0.0411_{-0.04}^{+0.0010}$ & $0.0416_{-0.021}^{+0.0008}$ \\
38 & $0.578_{-0.26}^{+0.011}$ & $0.0591_{-0.06}^{+0.0017}$ & $0.0595_{-0.0004}^{+0.0020}$ \\
39 & $0.5899_{-0.007}^{+0.0035}$ & $0.0797_{-0.0005}^{+0.0021}$ & $0.0802_{-0.0007}^{+0.0016}$ \\
40 & $0.574_{-0.21}^{+0.009}$ & $0.0878_{-0.09}^{+0.0033}$ & $0.0898_{-0.05}^{+0.0025}$ \\
41 & $0.595_{-0.28}^{+0.004}$ & $0.0954_{-0.09}^{+0.0011}$ & $0.0954_{-0.09}^{+0.0011}$ \\
42 & $0.567_{-0.21}^{+0.012}$ & $0.0378_{-0.04}^{+0.0011}$ & $0.0382_{-0.021}^{+0.0007}$ \\
43 & $0.589145_{-0.}^{+0.0017}$ & $0.0202925_{-0.}^{+0.00023}$ & $0.020302_{-0.}^{+0.0006}$ \\
44 & $0.593_{-0.016}^{+0.008}$ & $0.0517_{-0.0009}^{+0.0011}$ & $0.0535_{-0.0022}^{+0.0006}$ \\
45 & $0.595_{-0.21}^{+0.011}$ & $0.0825_{-0.08}^{+0.0008}$ & $0.0827_{-0.04}^{+0.0014}$ \\
46 & $0.597_{-0.020}^{+0.04}$ & $0.0513_{-0.013}^{+0.0014}$ & $0.0535_{-0.014}^{+0.0006}$ \\
47 & $0.426_{-0.004}^{+0.20}$ & $0.00000001_{-0.}^{+0.04}$ & $0.01872_{-0.00030}^{+0.020}$ \\
48 & $0.639_{-0.004}^{+0.007}$ & $0.0389_{-0.0005}^{+0.008}$ & $0.0389_{-0.0005}^{+0.008}$
    \end{tabular}
   \end{ruledtabular}
    \caption{Fit parameters for the triton.}
    \label{tab:fit_params_t}
\end{table}

\begin{table}[H]
        \begin{ruledtabular}
    \begin{tabular}{c ccc }
$t$ & $N$ & $\omega^+$ & $\omega^-$  \\ \hline 
1 & $2.49_{-0.15}^{+2.8}$ & $1.43_{-0.04}^{+1.3}$ & $\left(0.13_{-0.12}^{+6.}\right)\times 10^3$ \\
2 & $2.49_{-0.15}^{+5.}$ & $1.43_{-0.04}^{+2.5}$ & $\left(0.13_{-0.13}^{+5.}\right)\times 10^3$ \\
3 & $10.56_{-7.}^{+0.25}$ & $5.45_{-3.4}^{+0.12}$ & $\left(4.0_{-2.7}^{+8.}\right)\times 10^3$ \\
4 & $5.1_{-2.7}^{+2.2}$ & $2.7_{-1.3}^{+1.1}$ & $\left(0.15_{-0.15}^{+3.2}\right)\times 10^3$ \\
5 & $2.46_{-0.21}^{+2.7}$ & $1.42_{-0.07}^{+1.3}$ & $\left(0.009_{-0.006}^{+7.}\right)\times 10^3$ \\
6 & $3.61_{-0.9}^{+0.13}$ & $1.99_{-0.5}^{+0.07}$ & $\left(0.0055_{-0.0035}^{+1.2}\right)\times 10^5$ \\
7 & $2.807_{-0.07}^{+0.034}$ & $1.629_{-0.04}^{+0.020}$ & $\left(1.23_{-0.12}^{+1.2}\right)\times 10^4$ \\
8 & $2.34_{-0.12}^{+0.17}$ & $1.385_{-0.023}^{+0.05}$ & $\left(0.00175_{-0.00024}^{+3.2}\right)\times 10^3$ \\
9 & $1.863_{-0.11}^{+0.027}$ & $1.205_{-0.024}^{+0.031}$ & $\left(1.2_{-1.2}^{+0.7}\right)\times 10^4$ \\
10 & $1.680_{-0.024}^{+0.012}$ & $1.149_{-0.022}^{+0.019}$ & $3147._{-335.}^{+927.}$ \\
11 & $1.713_{-0.019}^{+0.024}$ & $1.233_{-0.023}^{+0.016}$ & $2652._{-418.}^{+501.}$ \\
12 & $1.509_{-0.017}^{+0.013}$ & $1.081_{-0.011}^{+0.013}$ & $11.22_{-0.23}^{+0.4}$ \\
13 & $1.393_{-0.012}^{+0.018}$ & $0.936_{-0.007}^{+0.021}$ & $4.87_{-0.10}^{+0.09}$ \\
14 & $1.284_{-0.008}^{+0.012}$ & $0.762_{-0.014}^{+0.013}$ & $2.396_{-0.015}^{+0.030}$ \\
15 & $1.200_{-0.012}^{+0.015}$ & $0.579_{-0.010}^{+0.009}$ & $1.263_{-0.014}^{+0.017}$ \\
16 & $1.148_{-0.027}^{+0.017}$ & $0.4057_{-0.0034}^{+0.010}$ & $0.685_{-0.009}^{+0.013}$ \\
17 & $1.126_{-0.017}^{+0.014}$ & $0.288_{-0.004}^{+0.007}$ & $0.398_{-0.008}^{+0.007}$ \\
18 & $1.132_{-0.005}^{+0.010}$ & $0.2000_{-0.0029}^{+0.0028}$ & $0.2419_{-0.0029}^{+0.005}$ \\
19 & $1.120_{-0.009}^{+0.018}$ & $0.1400_{-0.0019}^{+0.0035}$ & $0.1573_{-0.0034}^{+0.0023}$ \\
20 & $1.097_{-0.011}^{+0.009}$ & $0.0877_{-0.0006}^{+0.0018}$ & $0.0944_{-0.0007}^{+0.0010}$ \\
21 & $1.098_{-0.010}^{+0.026}$ & $0.05345_{-0.00032}^{+0.0018}$ & $0.0554_{-0.0010}^{+0.0018}$ \\
22 & $1.091_{-0.010}^{+0.014}$ & $0.03154_{-0.00024}^{+0.0007}$ & $0.03226_{-0.0006}^{+0.00024}$ \\
23 & $1.085_{-0.008}^{+0.013}$ & $0.01960_{-0.00010}^{+0.00027}$ & $0.01995_{-0.00010}^{+0.00016}$ \\
24 & $1.070_{-0.007}^{+0.016}$ & $0.01639_{-0.00014}^{+0.00020}$ & $0.01653_{-0.00019}^{+0.00024}$ \\
25 & $1.062_{-0.009}^{+0.006}$ & $0.00876_{-0.00013}^{+0.00011}$ & $0.00876_{-0.00006}^{+0.00011}$ \\
26 & $1.058_{-0.017}^{+0.025}$ & $0.001533_{-0.000017}^{+0.000034}$ & $0.001538_{-0.000022}^{+0.00004}$ \\
27 & $1.051_{-0.010}^{+0.023}$ & $0.00722_{-0.00020}^{+0.00011}$ & $0.00722_{-0.00016}^{+0.00011}$ \\
28 & $1.051_{-0.006}^{+0.022}$ & $0.00579_{-0.00006}^{+0.00012}$ & $0.00579_{-0.00004}^{+0.00012}$ \\
29 & $1.036_{-0.5}^{+0.022}$ & $0.01594_{-0.016}^{+0.00024}$ & $0.01594_{-0.010}^{+0.00030}$ \\
30 & $1.060_{-0.023}^{+0.013}$ & $0.01906_{-0.00031}^{+0.0004}$ & $0.01908_{-0.00025}^{+0.0004}$ \\
31 & $1.022_{-0.027}^{+0.007}$ & $0.01265_{-0.00028}^{+0.00025}$ & $0.01277_{-0.00033}^{+0.00016}$ \\
32 & $1.027_{-0.009}^{+0.18}$ & $0.000574_{-0.000017}^{+0.00012}$ & $0.000580_{-0.000018}^{+0.00013}$ \\
33 & $1.014_{-0.033}^{+0.021}$ & $0.00528_{-0.00016}^{+0.00016}$ & $0.00528_{-0.00016}^{+0.00016}$ \\
34 & $1.017_{-0.011}^{+0.023}$ & $0.000665_{-0.000007}^{+0.000012}$ & $0.000665_{-0.000007}^{+0.000012}$ \\
35 & $0.996_{-0.5}^{+0.019}$ & $0.00941_{-0.009}^{+0.00019}$ & $0.00946_{-0.006}^{+0.00016}$ \\
36 & $1.0006_{-0.022}^{+0.0028}$ & $0.01138_{-0.00028}^{+0.00011}$ & $0.01146_{-0.00032}^{+0.00013}$ \\
37 & $0.991_{-0.012}^{+0.007}$ & $0.01441_{-0.00017}^{+0.00024}$ & $0.01457_{-0.00016}^{+0.00034}$ \\
38 & $0.981_{-0.007}^{+0.009}$ & $0.002511_{-0.000032}^{+0.000028}$ & $0.002536_{-0.000005}^{+0.000032}$ \\
39 & $0.9635_{-0.0026}^{+0.012}$ & $0.001160_{-0.000016}^{+0.000014}$ & $0.001164_{-0.000018}^{+0.000011}$ \\
40 & $0.968_{-0.4}^{+0.010}$ & $0.00654_{-0.006}^{+0.00004}$ & $0.006549_{-0.004}^{+0.000029}$ \\
41 & $0.951_{-0.006}^{+0.012}$ & $0.00824_{-0.00007}^{+0.00025}$ & $0.00831_{-0.00009}^{+0.00020}$ \\
42 & $0.939_{-0.015}^{+0.006}$ & $0.02217_{-0.00031}^{+0.00028}$ & $0.02217_{-0.00018}^{+0.00028}$ \\
43 & $0.945_{-0.007}^{+0.012}$ & $0.02977_{-0.00031}^{+0.0004}$ & $0.03044_{-0.0004}^{+0.00018}$ \\
44 & $0.939_{-0.4}^{+0.012}$ & $0.0315_{-0.031}^{+0.0009}$ & $0.0327_{-0.020}^{+0.0007}$ \\
45 & $0.963_{-0.006}^{+0.014}$ & $0.01731_{-0.00022}^{+0.00019}$ & $0.01740_{-0.00030}^{+0.0004}$ \\
46 & $0.966_{-0.011}^{+0.017}$ & $0.00950_{-0.00014}^{+0.00029}$ & $0.00973_{-0.00019}^{+0.00013}$ \\
47 & $1.009_{-0.4}^{+0.018}$ & $0.001032_{-0.0006}^{+0.000005}$ & $0.001046_{-0.0006}^{+0.000016}$ \\
48 & $1.053_{-0.04}^{+0.018}$ & $0.001261_{-0.00006}^{+0.000035}$ & $0.001268_{-0.00005}^{+0.000031}$
    \end{tabular}
   \end{ruledtabular}
    \caption{Fit parameters for the $\alpha$.}
    \label{tab:fit_params_a}
\end{table}

\newpage
\begin{figure*}
    \centering
    \includegraphics[width=0.83\textwidth]{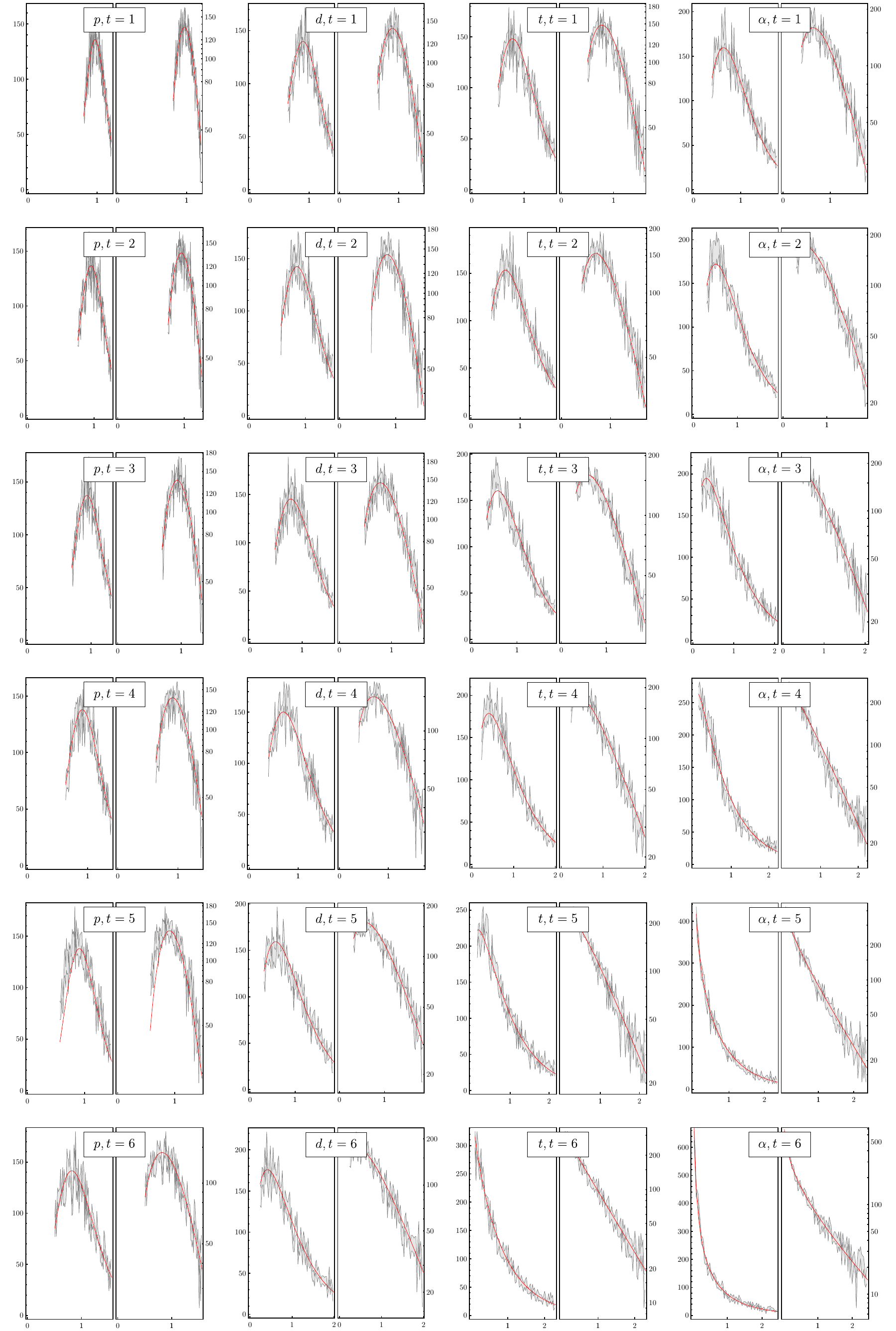}
    \caption{Correlation function distributions for $t\in\{1,\ldots,6\}$ and $B\in\{p,d,t,\alpha\}$. Grey regions correspond to the empirical histograms and red curves correspond to the best fit using Eq.~\eqref{eq:shifted_func}.
    \label{fig:dist_1}}
\end{figure*}
\begin{figure*}
    \centering
    \includegraphics[width=0.83\textwidth]{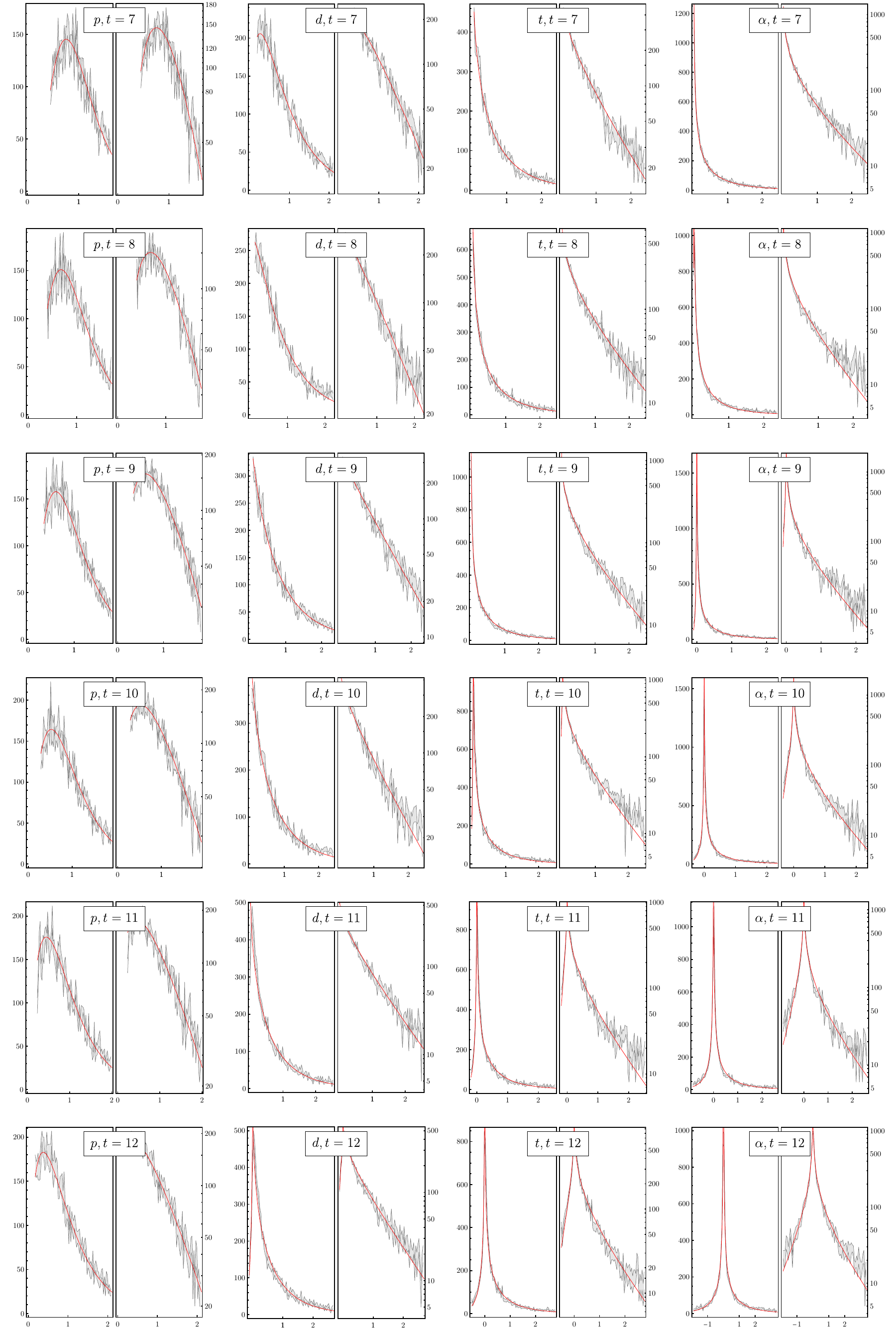}
    \caption{Correlation function distributions for $t\in\{7,\ldots,12\}$ and $B\in\{p,d,t,\alpha\}$. Same conventions as in Fig.~\ref{fig:dist_1}.}
\end{figure*}
\begin{figure*}
    \centering
    \includegraphics[width=0.83\textwidth]{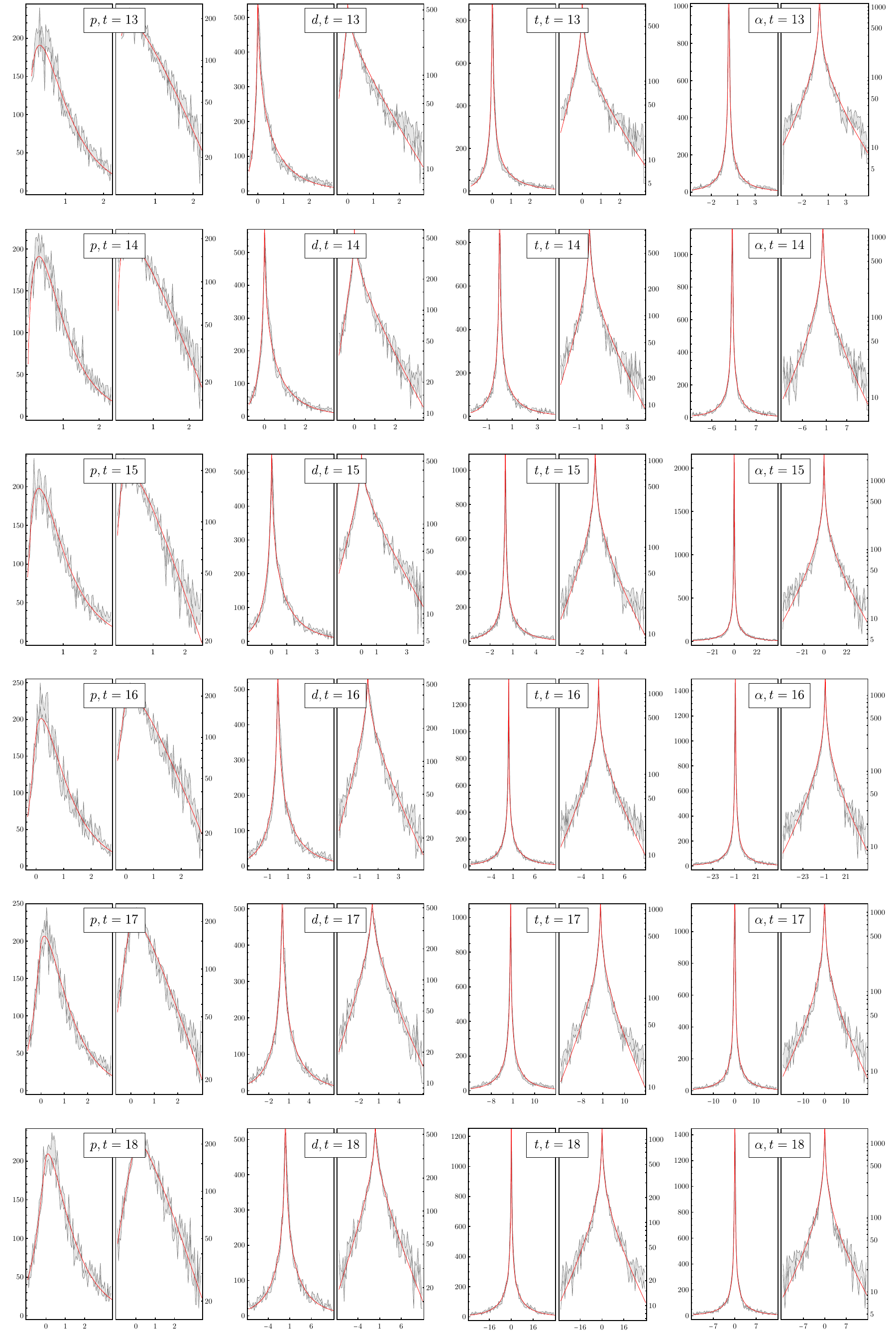}
    \caption{Correlation function distributions for $t\in\{13,\ldots,18\}$ and $B\in\{p,d,t,\alpha\}$. Same conventions as in Fig.~\ref{fig:dist_1}.}
\end{figure*}
\begin{figure*}
    \centering
    \includegraphics[width=0.83\textwidth]{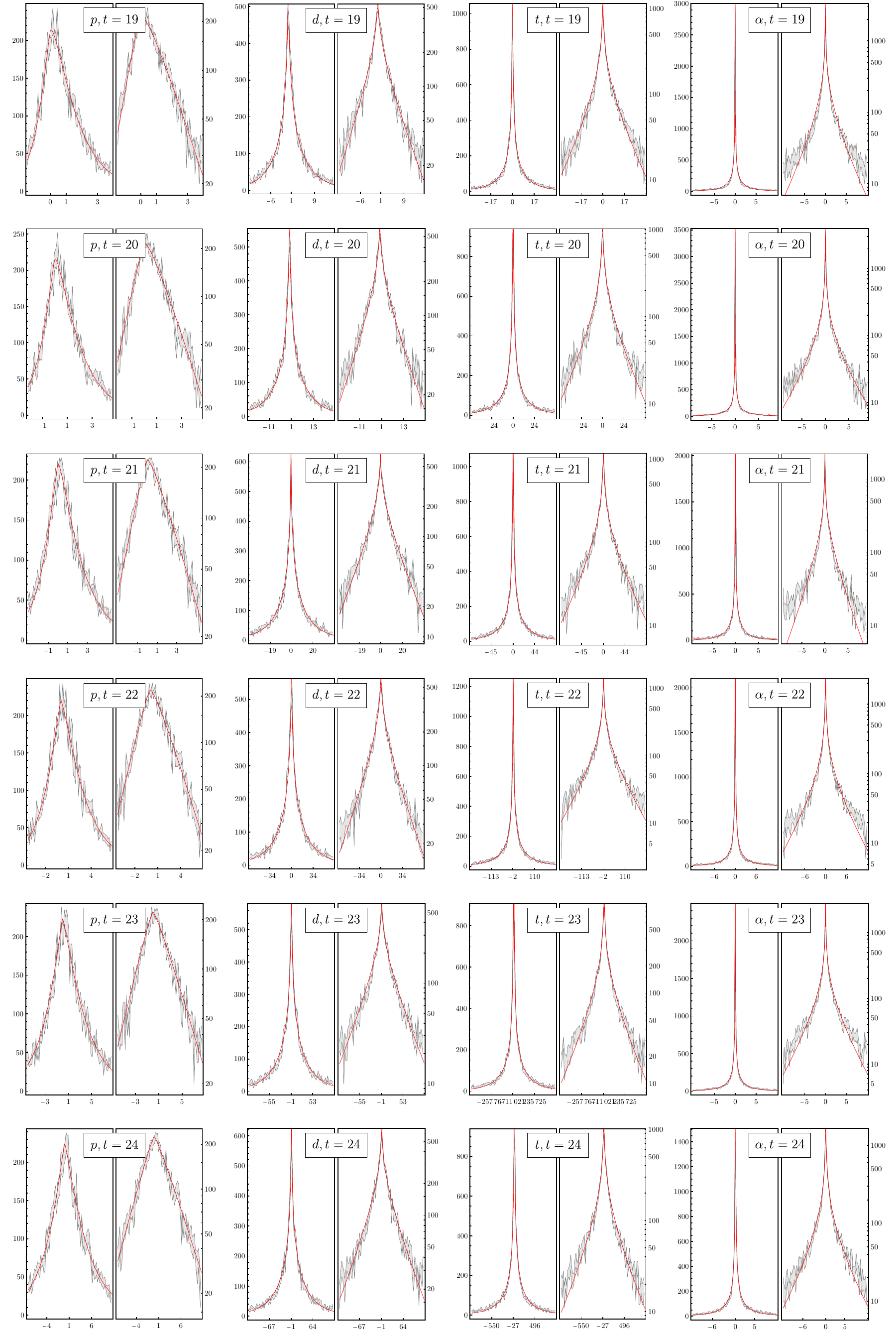}
    \caption{Correlation function distributions for $t\in\{19,\ldots,24\}$ and $B\in\{p,d,t,\alpha\}$. Same conventions as in Fig.~\ref{fig:dist_1}.}
\end{figure*}
\begin{figure*}
    \centering
    \includegraphics[width=0.83\textwidth]{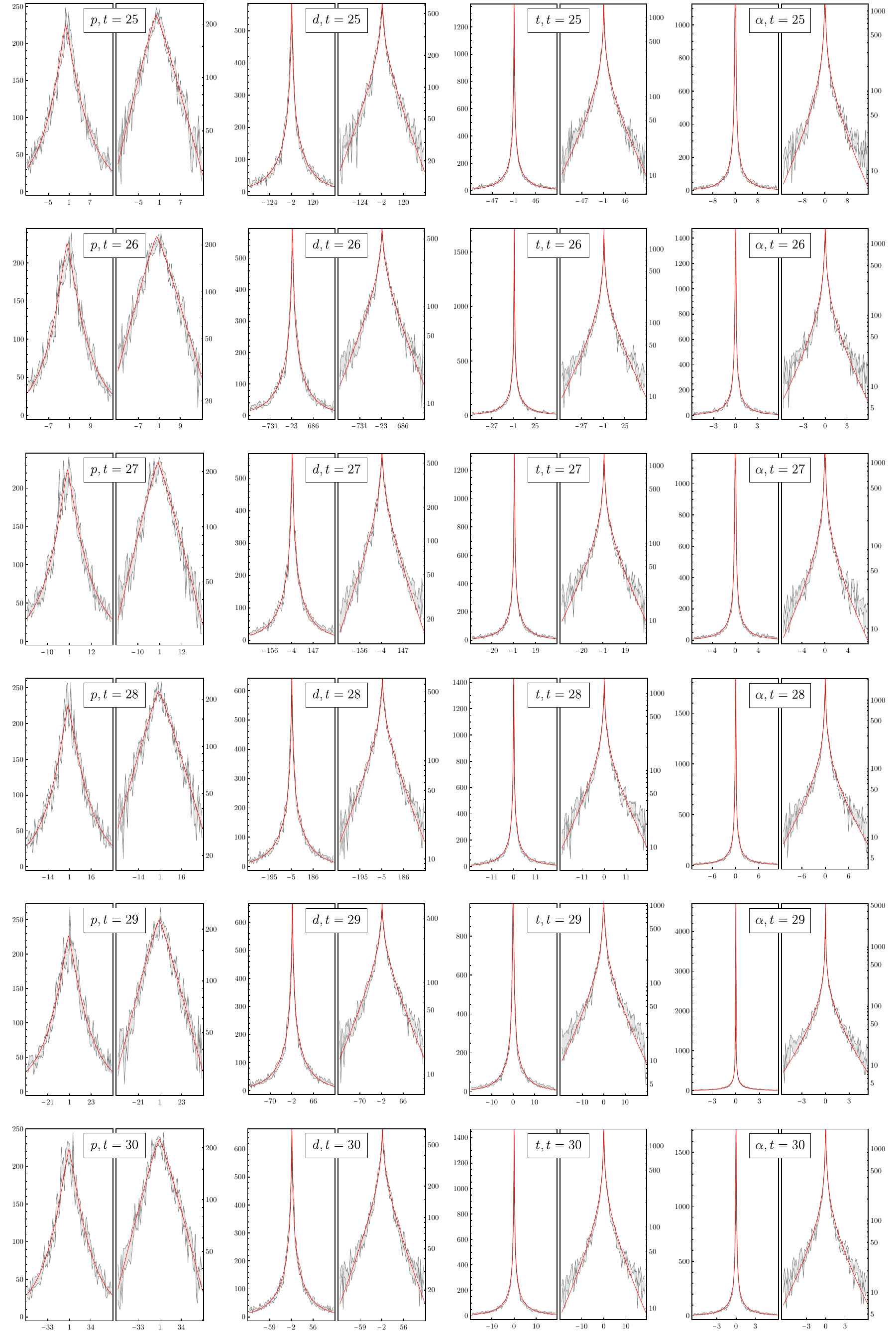}
    \caption{Correlation function distributions for $t\in\{25,\ldots,30\}$ and $B\in\{p,d,t,\alpha\}$. Same conventions as in Fig.~\ref{fig:dist_1}.}
\end{figure*}
\begin{figure*}
    \centering
    \includegraphics[width=0.83\textwidth]{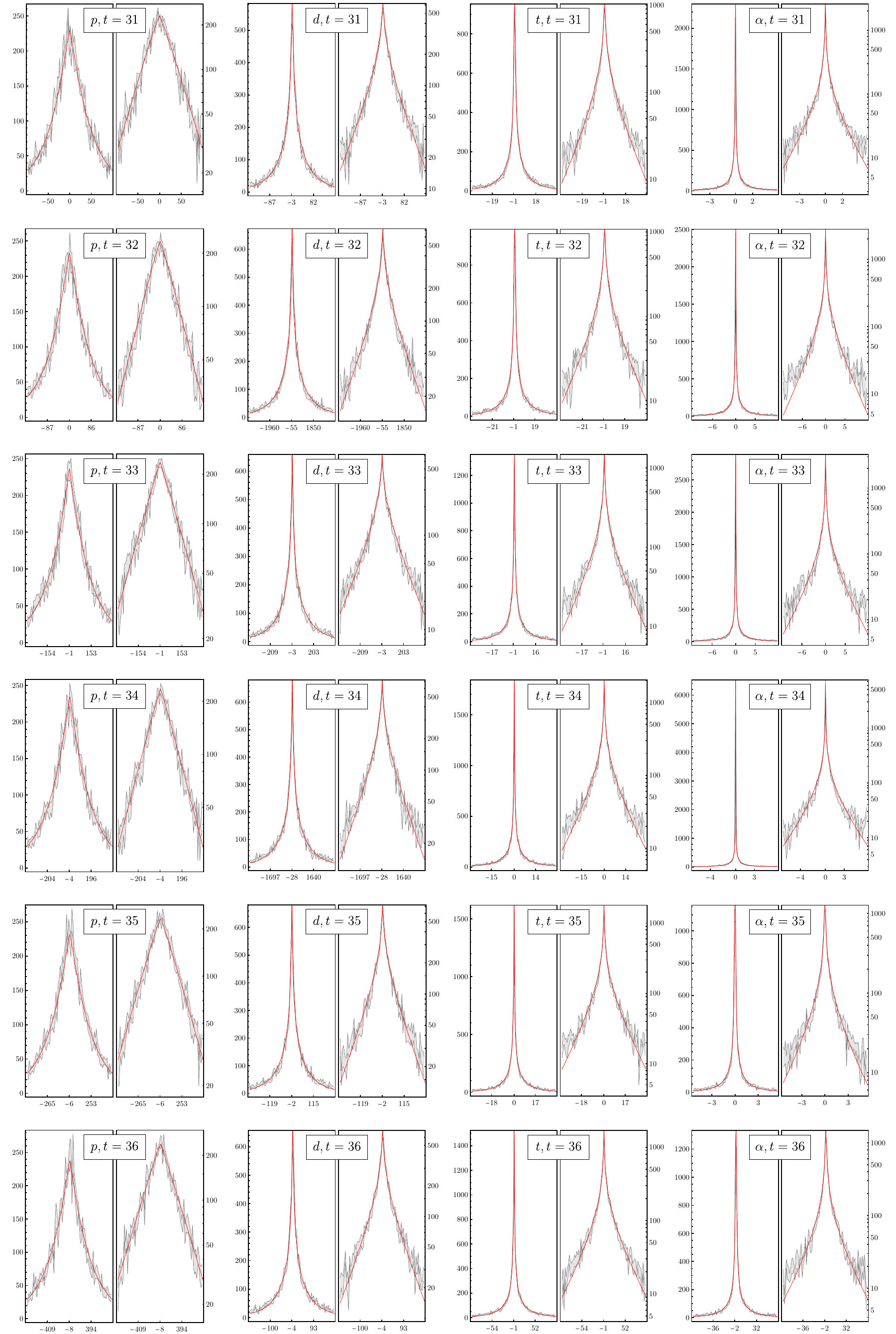}
\caption{Correlation function distributions for $t\in\{31,\ldots,36\}$ and $B\in\{p,d,t,\alpha\}$. Same conventions as in Fig.~\ref{fig:dist_1}.}
\end{figure*}
\begin{figure*}
    \centering
    \includegraphics[width=0.83\textwidth]{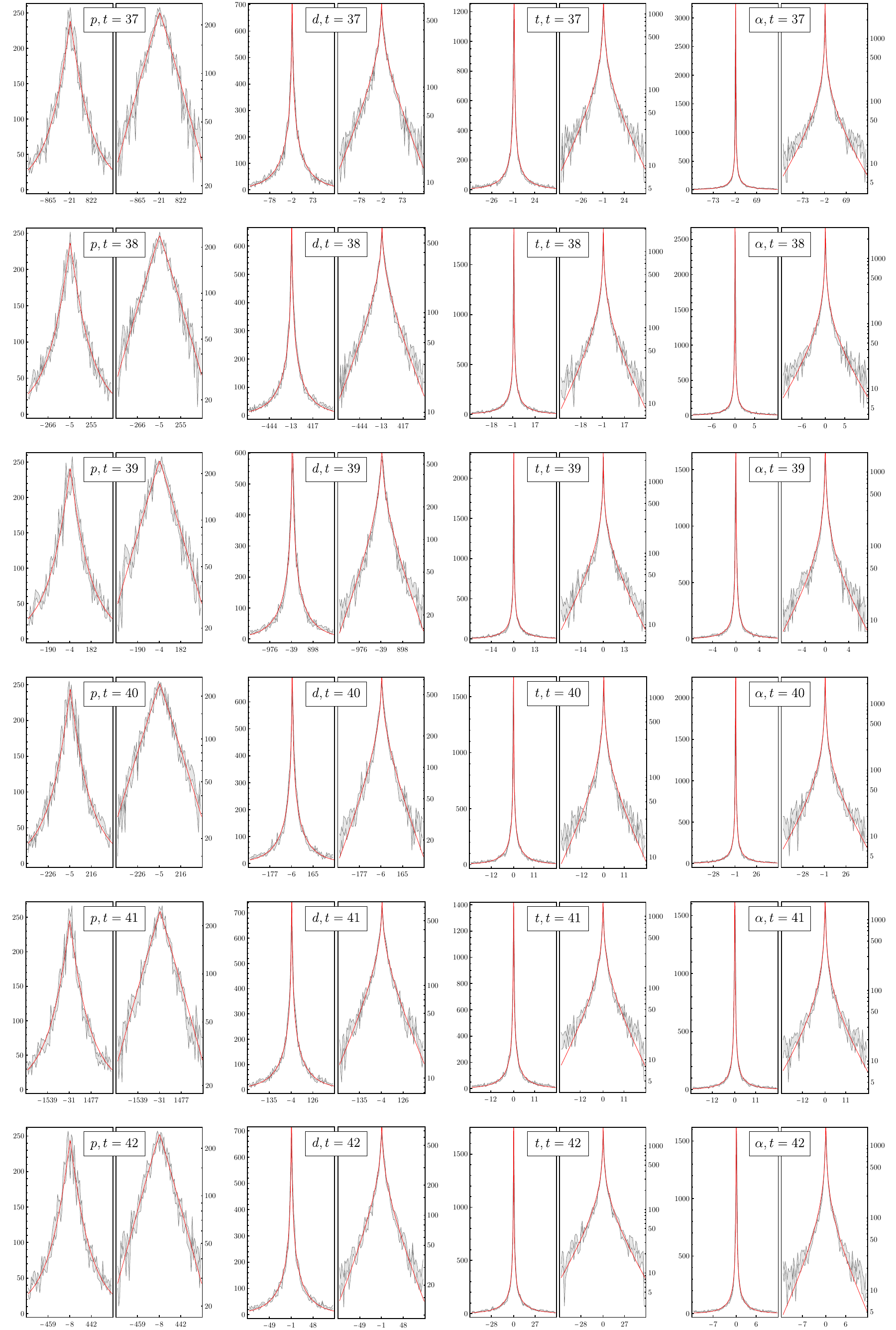}
    \caption{Correlation function distributions for $t\in\{37,\ldots,42\}$ and $B\in\{p,d,t,\alpha\}$. Same conventions as in Fig.~\ref{fig:dist_1}.}
\end{figure*}
\begin{figure*}
    \centering
    \includegraphics[width=0.83\textwidth]{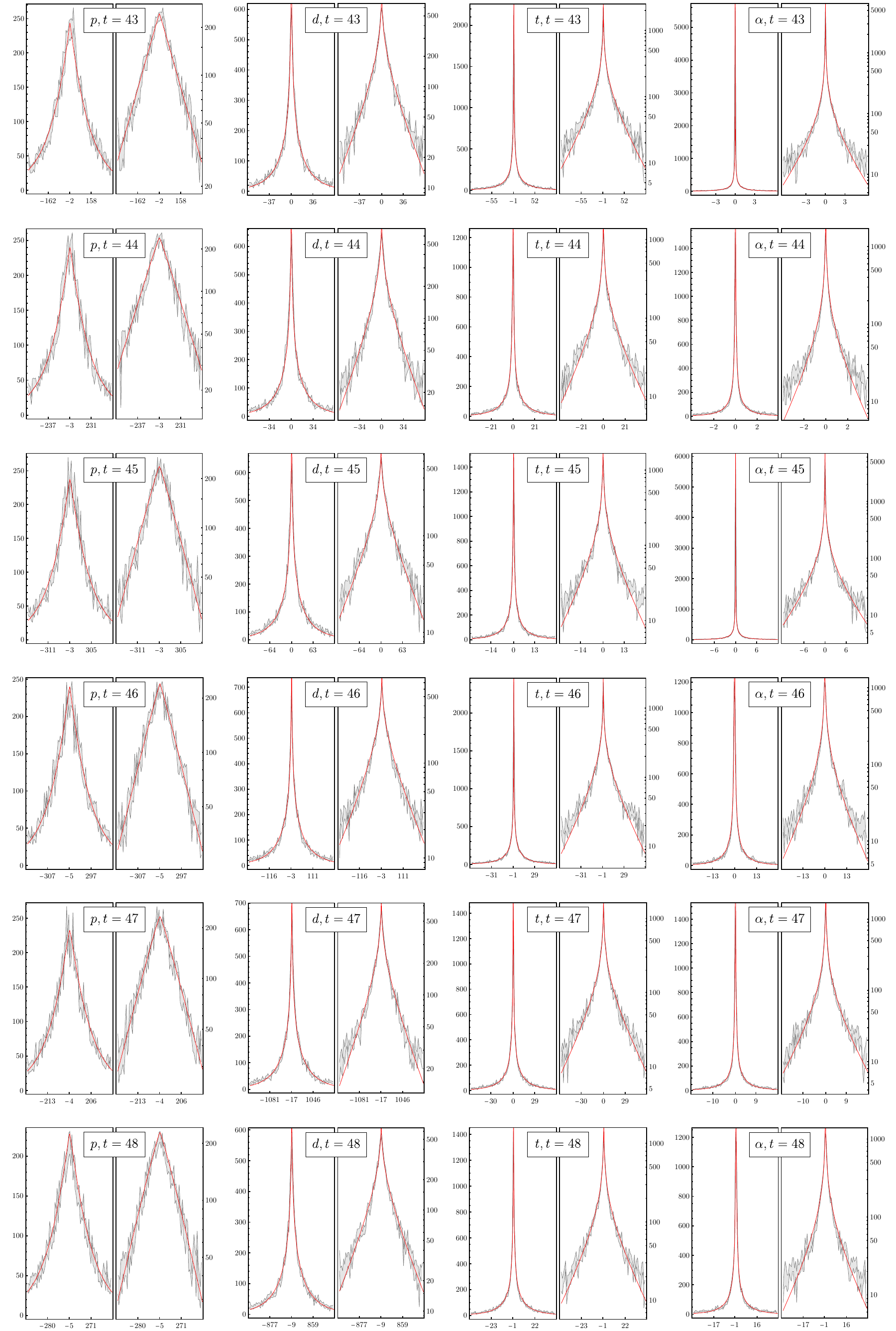}
    \caption{Correlation function distributions for $t\in\{43,\ldots,48\}$ and $B\in\{p,d,t,\alpha\}$. Same conventions as in Fig.~\ref{fig:dist_1}.    \label{fig:dist_8}}
\end{figure*}

\end{document}